%% file: main.tex
\documentclass[a4paper,fleqn]{cas-sc}

\usepackage[numbers,sort&compress]{natbib}

\def\figureautorefname~#1\null{Fig.\,#1\null}
\def\tableautorefname~#1\null{Tab.\,#1\null}

\def\equationautorefname~#1\null{Eq.\,(#1)\null}

\graphicspath{ {./plots/} }

\newlength{\dhatheight}

\include{macros}

\begin{document}
\let\WriteBookmarks\relax
\def\floatpagepagefraction{1}
\def\textpagefraction{.001}

\shorttitle{{\tt EvoEMD}}

\shortauthors{Dutra and Wu}

\title [mode = title]{{\tt EvoEMD}: cosmic evolution with an early matter-dominated era}

\author[1]{Ma\'ira Dutra}[orcid=0000-0003-4851-242X]
\ead{mdutra@physics.carleton.ca}
\affiliation[1]{organization={Ottawa-Carleton Institute for Physics,
Carleton University},
            addressline={1125 Colonel By Drive},
            city={Ottawa},
            postcode={K1S 5B6},
            state={ON},
            country={Canada}}

\author[2,3]{Yongcheng Wu}[orcid=0000-0002-1835-7660]
\cormark[2]
\ead{ycwu@njnu.edu.cn}
\affiliation[2]{organization={Department of Physics and Institute of Theoretical Physics, Nanjing Normal University},
city={Nanjing},
postcode={210023},
country={China}}
\affiliation[3]{organization={Department of Physics, Oklahoma State University},
            city={Stillwater},
            postcode={74078},
            state={OK},
            country={USA}}

\cortext[2]{Corresponding author}


\begin{abstract}
We present {\tt EvoEMD}, a framework to calculate the evolution of cosmic relics in a Universe with an early matter-dominated (EMD) era. There are mainly two aspects to consider in this regard. First, an EMD era changes the Hubble expansion rate with respect to the standard radiation-dominated (RD) universe. Second, when the EMD era ends, the out-of-equilibrium decay of the dominant matter component may reheat the thermal bath and dilute cosmic relics. We
briefly introduce the cosmology with an EMD era, and present how it is implemented in the {\tt EvoEMD} framework. Users can study the coupled evolution of different interacting species in an EMD or RD universe. Two important cosmic relics are dark matter and a net lepton number. In order to show the capabilities of {\tt EvoEMD}, we include simple examples of dark matter produced via freeze-out and freeze-in, and also of leptogenesis. Moreover, users can modify the model files in order to explore different new physics scenarios. {\tt EvoEMD} is hosted on {\tt Github} at \url{https://github.com/ycwu1030/EvoEMD}.
\end{abstract}




\maketitle

\flushbottom

\section{Introduction}

Our current scientific understanding of the universe and its fundamental components relies on the standard cosmological model ($\Lambda$CDM) and the standard model of particle physics (SM). $\Lambda$CDM successfully describes the observed nearly flat, homogeneous, and isotropic expanding universe containing a cosmological constant $\Lambda$ and cold dark matter (CDM), with gravity described by general relativity \cite{Planck:2018vyg}. It also encompasses the inflationary paradigm regarding the initial conditions for the observed universe. On the other hand, the SM successfully describes the fundamental particles and non-gravitational interactions at an astonishing precision \cite{ParticleDataGroup:2020ssz}.

However, both standard models face serious theoretical and observational challenges (see \cite{Bull:2015stt,Perivolaropoulos:2021jda} and \cite{Lee:2019zbu} for recent reviews). It is interesting to note that in order to deal with the successful predictions and also with the problems of $\Lambda$CDM, we must consider beyond the SM (BSM) scenarios. For instance, cosmic relics like dark matter (DM) particles, massive neutrinos, and the matter-antimatter asymmetry, are all in the BSM context. The origins of such relics are subject to intense research.

From the measurements of the cosmic microwave background (CMB), we know that ordinary matter (photons, electrons, etc) were part of a thermal bath, at a temperature of $T\sim 0.2$ eV. In the context of $\Lambda$CDM, the universe is at first dominated by the vacuum energy of the field driving inflation, the inflaton. After inflation, the out-of-equilibrium decay of the inflaton produces SM fields, and therefore the cosmic entropy. At the so-defined reheat temperature $T_{RH}$, a thermal bath of ultra-relativistic species (or radiation) is established, and the universe becomes \textit{radiation-dominated} (RD). The scale of the inflationary reheating $T_{RH}$ is not known, but it must be above the MeV scale in order to not spoil the BBN predictions \cite{Hannestad:2004px,Jedamzik:2006xz,Kawasaki:2017bqm}. Therefore, within $\Lambda$CDM the universe was RD from $T_{RH}$ up to $T\sim 0.75$ eV, when it becomes \textit{matter-dominated} (MD) due to the CDM component. Nowadays, the $\Lambda$ component dominates the total energy density and drives the accelerated cosmic expansion.

Matter-dominated periods are actually a direct consequence of BSM physics. On one hand, the SM does not include viable CDM candidates leading to the standard MD era. On the other hand, when a field which is part of the thermal bath (as all SM fields) becomes non-relativistic, behaving as matter, its energy density is exponentially suppressed and stops contributing to the total energy density of the universe. Decoupled and long-lived BSM fields can lead to \textit{early matter-dominated} (EMD) periods prior to BBN in many well-motivated models \cite{Adhya:2003wj,Asaka:2006ek,Hasenkamp:2010if,Co:2015pka,Berlin:2016vnh,Dutra:2021lto,Ertas:2021xeh}. For sufficiently long-lived matter components, EMD periods are followed by reheating periods (also required to end prior to BBN), in which a significant amount of entropy is injected into the SM bath \cite{Scherrer:1984fd}. Such a possibility is currently unconstrained but might have detectable signatures \cite{Erickcek:2011us,StenDelos:2019xdk,Barenboim:2021swl}.

EMD eras can significantly impact the phenomenology of cosmic relics. Any pre-existing DM relic or asymmetry is diluted if the entropy produced after the EMD era is large. DM candidates which were once thermalized with the SM thermal bath, becoming a relic via the \textit{freeze-out} mechanism, usually need to be more weakly interacting in order to compensate the dilution \cite{Hamdan:2017psw}. As a consequence, frozen-out relics can evade the current strong bounds on their couplings to SM particles without overclosing the universe. Moreover, EMD periods enable frozen-out relics beyond the upper unitarity limit on their masses \cite{Cirelli:2018iax,Bhatia:2020itt,Asadi:2021bxp,Bian:2021vmi}. On the other hand, DM particles which have never been thermalized with the SM bath, produced via the \textit{freeze-in} mechanism, need to interact more strongly in order to compensate the dilution. Therefore, EMD periods happening after freeze-in increases the testability of any frozen-in DM candidate \cite{Co:2015pka,Cosme:2020mck}.

There are many public codes dedicated to the study of cosmic relics. The dark matter relic density can be accurately computed with {\tt micrOMEGAs} \cite{Belanger:2006is,Belanger:2018ccd}, {\tt DarkSUSY} \cite{Bringmann:2018lay}, {\tt madDM} \cite{Ambrogi:2018jqj}, and {\tt DRAKE} \cite{Binder:2021bmg}. However,
these codes assume the standard case in which the universe was RD during freeze-out/freeze-in.
The net lepton number which can lead to the matter-antimatter asymmetry can be studied with {\tt ULYSSES} \cite{Granelli:2020pim} in standard or non-standard cosmological scenarios (see Ref. \cite{Perez-Gonzalez:2020vnz} for the study of leptogenesis during primordial black hole domination with {\tt ULYSSES}).

In this work we present {\tt EvoEMD}, a {\tt C++} framework for studying the evolution of cosmic relics in the presence of an EMD era. The duration of the EMD era can be controlled by users, so they can also consider the case of a standard RD universe. {\tt EvoEMD} allows the users to solve coupled Boltzmann fluid equations in BSM models which can be implemented in a relatively straightforward manner. In order to show the capabilities of {\tt EvoEMD}, we consider the freeze-out and the freeze-in production of dark matter and also leptogenesis in simple BSM models.

The structure of this paper is as follow. In \autoref{sec:EMD}, we discuss how the cosmic expansion is affected by an EMD era. In \autoref{sec:BE}, we present the Boltzmann fluid equation governing the evolution of interacting species taking into account the possibility of an EMD and a late reheating periods. The structure of {\tt EvoEMD} is detailed in \autoref{sec:EvoEMD}, and examples of how to use the code to study DM and leptogenesis are shown in \autoref{sec:Examples}. Finally, we summarize in \autoref{sec:Summary}.

\section{Early Matter-Dominated Era}
\label{sec:EMD}

In this section, we review some aspects of an early matter-dominated era.
The universe expands at a rate which depends on its energetic constituents. In an homogeneous, isotropic, and flat universe, the expansion is governed by the Friedmann equation:
\begin{equation}
    H(t)^2 = \frac{8\pi G}{3}\rho(t) \equiv \frac{\rho(t)}{3 M_P^2} \,,
\end{equation}
with $H(t)\equiv \dot{a}/a$ the Hubble expansion rate, $a(t)$ the time-dependent scale factor, $\rho(t)$  the total energy density, $G$ the Newton's constant, and $M_P \simeq 2.43 \times 10^{18}$ GeV the reduced Planck mass.

In the absence of collisions and heat flows, energy conservation in an expanding universe implies that the energy density of a given species $i$, $\rho_i$, redshifts as $\rho_i = \rho_i^0 \left(\frac{a}{a_0}\right)^{-3(1+w_i)}$, with $w_i$ parameterizing its equation of state ${\cal P}_i = w_i \rho_i$\footnote{If the component is a vacuum energy, we have $w_i = -1$; if it is radiation, $w_i=1/3$; if it is matter, $w_i=0$.}. When the total energy is dominated by one of the cosmic components, we say that the cosmic expansion is dominated by that component.

\begin{figure}[!tb]
    \centering
    \includegraphics[width=0.6\textwidth]{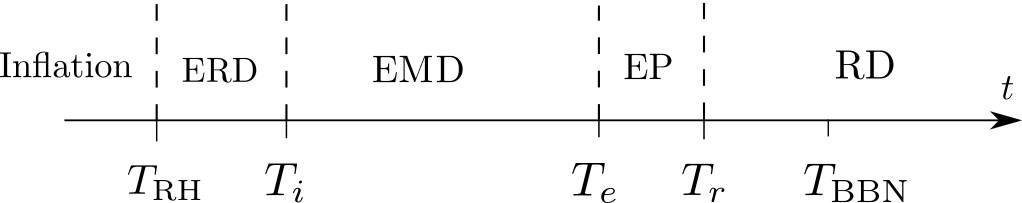}
    \caption{The evolution of the Universe with an EMD era.}
    \label{fig:evo_emd}
\end{figure}

The cosmic history in the non-standard context of an early matter domination can be briefly summarized in~\autoref{fig:evo_emd}. After the inflationary reheating at $T_{RH}$ the standard radiation-dominated (RD) era is interrupt such that we have an early RD (ERD) era followed by an isentropic early matter-dominated (EMD) era which starts at an initial temperature $T_i$ and ends at a temperature $T_e$. The out-of-equilibrium decay of the matter field driving the EMD era into SM fields leads to a period of entropy production (EP), reheating the thermal bath. The EP period finishes at a late reheat temperature $T_r$, which must be above the BBN scale $T_{BBN} \sim$ MeV.

The heat flow due to the decay of the decoupled matter field into the radiation component couples the evolution of their energy densities, respectively $\rho_M$ and $\rho_R$. This is governed by the following Boltzmann equations:
\begin{subequations}
\begin{align}
    \frac{d\rho_M}{dt} + 3H\rho_M &= -\rho_M\Gamma_M \\
    \frac{d\rho_R}{dt} + 4H\rho_R &= f\rho_M\Gamma_M \,,
\end{align}
\end{subequations}
where
$\Gamma_M$ is the total decay width of the matter component and $f$ is the branching fraction of the matter component decaying into the radiation component. Note that if the matter field decays into another component besides radiation ($f\neq 1$), one must consider the evolution of the energy density for that component. At the relevant temperatures ($T_{BBN}< T< T_{RH}$) we assume $H=\sqrt{\frac{\rho_M+\rho_R}{3M_P^2}}$.

Using the definition of the Hubble rate ($1/dt = aH/da$), we can recast the equations above in the form
\begin{subequations}
    \label{equ:BE_rho}
    \begin{align}
        \frac{d(\rho_M a^3)}{da} &= -\frac{a^2 \rho_M \Gamma_M}{H}\,,\\
        \frac{d(\rho_R a^4)}{da} &= f\frac{a^3 \rho_M \Gamma_M}{H}\,.
    \end{align}
\end{subequations}
We can further simplify these equations by defining dimensionless quantities. We re-scale the scale factor by $x \equiv k a$ and the energy densities by the quantities
\begin{subequations}
\begin{align}
    Y_1 &= \rho_M a^3 \frac{k^3}{3M_P^2\Gamma_M^2}  = \frac{\rho_M x^3}{3M_P^2\Gamma_M^2}\,,\\
    Y_2 &= \rho_R a^4 \frac{k^4}{3M_P^2\Gamma_M^2} = \frac{\rho_R x^4}{3M_P^2\Gamma_M^2}\,,
\end{align}
\end{subequations}
which are comoving in the absence of heat flows and absorb most of the physical parameters. Note that the factor $k$ does not affect physical quantities but is used to conveniently shift $a$.

In terms of $x,Y_1$, and $Y_2$, the Boltzmann equations become
\begin{subequations}\label{equ:evolution}
    \begin{align}
        \frac{dY_1}{dx} &= -\frac{x Y_1}{\sqrt{x Y_1 + Y_2}}\,,\\
        \frac{dY_2}{dx} &= f\frac{x^2 Y_1}{\sqrt{x Y_1 + Y_2}} \,.
    \end{align}
\end{subequations}

To provide the initial conditions for $Y_{1,2}$, we define $T_i$ and $T_r$ as the temperatures at which $\rho_M = \rho_R$ and assume that $x(T_i) = 1$. On the other hand, the decay of the matter component is prominent when the decay width is roughly the Hubble parameter at that temperature. Then we assume $\Gamma_M = \kappa H(T_r)$, where $\kappa$ will be determined from the definition of $\rho_M(T_r) = \rho_R(T_r)$ and will be described in the following. Then the initial conditions are given by
\begin{subequations}
\label{equ:rhoM_rhoR_initial}
\begin{align}
Y_1|_{x=1} = \frac{\rho_R(T_i)}{\kappa^2\rho_R(T_r)}\,,\\
Y_2|_{x=1} = \frac{\rho_R(T_i)}{\kappa^2\rho_R(T_r)}\,.
\end{align}
\end{subequations}

With the above initial conditions and given $\kappa$ as well as the branch fraction $f$, the evolution of $Y_1$ ($\rho_M$) and $Y_2$ ($\rho_R$), and hence the SM bath temperature $T \equiv \left(\frac{\pi^2}{30}\frac{g_e(T)}{\rho_R}\right)^{-1/4} $, with $x$ ($a$) can be solved numerically according to~\autoref{equ:evolution}. Then, $T_i$ and $T_r$ can be determined by their definitions $\rho_M=\rho_R$. Note that with the above initial conditions, $T_i$ automatically matches its input. However, in general, $T_r$ will be shifted from its input. Then $\kappa$ is tuned to have $T_r$ matched.

\begin{figure}[!tb]
    \centering
    \includegraphics[width=.9\textwidth]{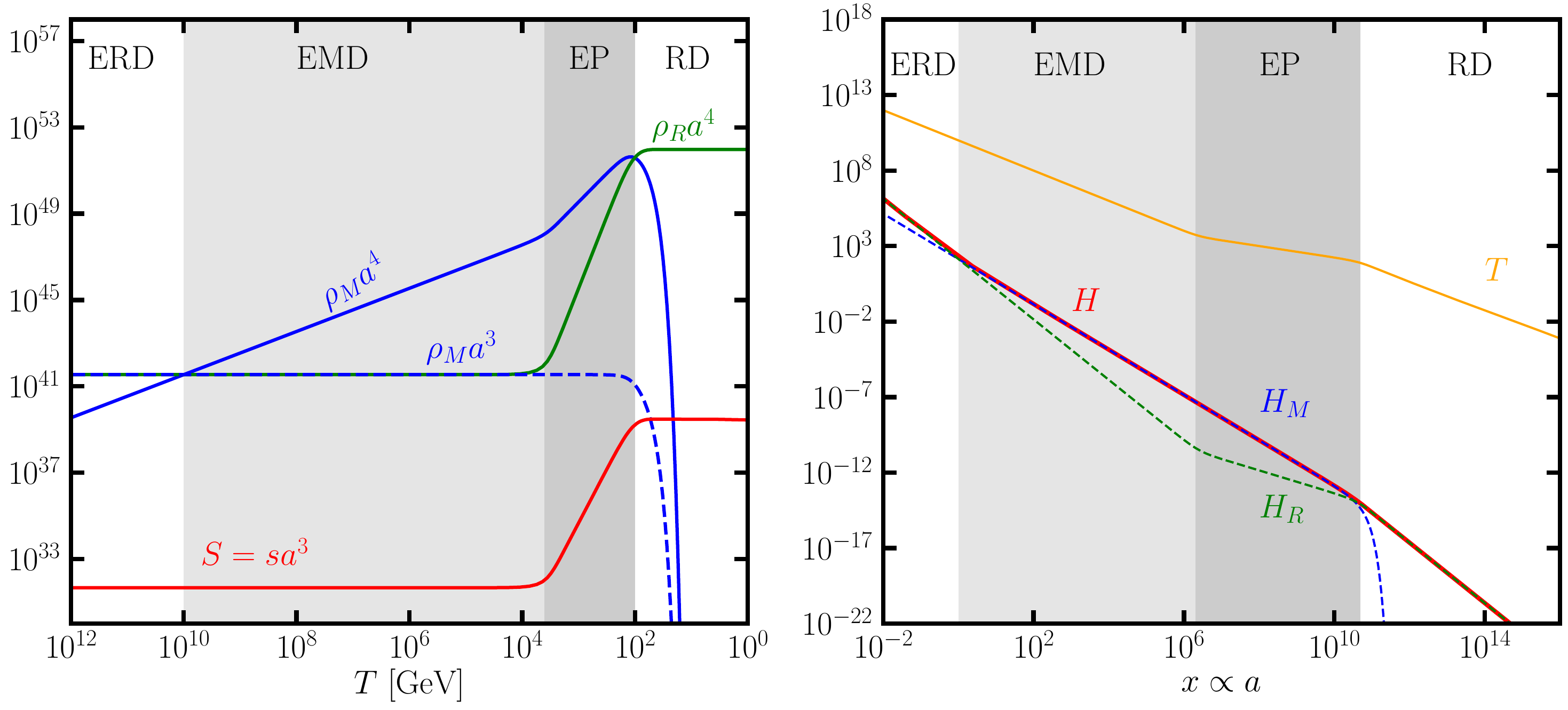}
    \caption{The evolution of relevant physical quantities. \textit{Left}: Comoving energy densities for radiation ($\rho_R a^4$, green curve) and the extra matter ($\rho_M a^3$, dashed blue curve), as well as $\rho_M a^4$ (solid blue curve) and total entropy ($S$, red curve). \textit{Right}: The resulting temperature of the thermal bath ($T$, orange curve) and the total Hubble parameter ($H$, red curve), together with the contributions of the radiation ($H_R$, dashed green curve) and the matter ($H_M$, dashed blue curve) components. As we can see, the transition between the different eras can be considered as ``instantaneous''.
    }
    \label{fig:rho_s_T_H_evo}
\end{figure}

In \autoref{fig:rho_s_T_H_evo} we show the solution of \autoref{equ:evolution} for $T_i=10^{10}$ GeV, $T_r=10^2$ GeV and $f=1$. In the left panel, we show, up to some normalization factor determined by fixing $a=1$ at today, the comoving energy densities of matter (dashed blue curve) and radiation (solid green curve) and the quantity $\rho_M a^4$ (solid blue curve) as functions of temperature. As we can see, we have at first an ERD era ($\rho_R > \rho_M$). When $\rho_M > \rho_R$, an isentropic EMD era starts. When the decay of the matter component into radiation becomes efficient ($\Gamma_M \sim H$), we have an EP period in which the total entropy in a comoving volume, $S$, increases due to the increase in $\rho_R$. We also show the solution for the entropy (red curve), which obeys $\dot S = \rho_M a^3 \Gamma_M/T$. In the right panel, we see the temperature of the thermal bath (orange curve) and the Hubble rate (red curve) as functions of the re-scaled scale factor $x = k a$. We can clearly see the transition between each era throughout evolution: during the ERD era the Hubble is dominated by radiation ($H_R$, dashed green curve), then by matter in the EMD and EP eras ($H_M$, dashed blue curve), and finally by radiation again in the RD era. During the EP era, though, the relation between temperature and scale factor is different, and so the temperature-dependence of the Hubble rate.

\subsection{Hubble parameter}
\label{sec:Hubble}

From our previous results, we conclude that it is reasonable to treat the transition between the different eras as ``instantaneous", which greatly simplifies our numerical study. In what follows, we derive the relation between the Hubble parameter and the temperature by splitting into different eras.

\subsubsection*{For the ERD/RD era}
During the radiation-dominated era (either ERD or RD), we have
\begin{align}
    H_{ERD/RD} = \sqrt{\frac{\rho_R}{3M_P^2}} = \frac{\pi}{3}\sqrt{\frac{g_e(T)}{10}}\frac{T^2}{M_P},
\end{align}
where we used the definition of the energy density of the radiation $\rho_R = \frac{\pi^2}{30}g_e(T)T^4$, with $g_e(T)$ the energetic relativistic degrees of freedom at $T$. During either ERD or RD, there is no extra source producing radiation, then we have $\rho_R\propto a^{-4}\propto g_e(T)T^4$.

\subsubsection*{For the EMD era}
During the initial stage of the matter-dominated era, the heat flow due to the decay of the matter component into radiation is not yet efficient.
Then we have:
\begin{subequations}
\begin{align}
    \rho_M a^3 &= {\rm Const}\\
    S = sa^3 &= {\rm Const} \,,
\end{align}
\end{subequations}
with $s = \frac{\pi^2}{45}g_s(T)T^3$ the entropy density and $g_s(T)$ the entropic relativistic degrees of freedom at $T$.

At the beginning of EMD era, $T=T_i$, the energy density of the matter component is equal to that of the radiation component, $\rho_M^i = \rho_R^i = \frac{\pi^2}{30}g_e(T_i)T_i^4$. During the EMD era, while the matter component decay is negligible,
we have
\begin{align}
    \rho_M &= \rho_M^i \left(\frac{a_i}{a}\right)^3 = \rho_M^i \frac{s}{s_i}\nonumber\\
    & = \rho_R^i \frac{g_s(T)}{g_s(T_i)}\frac{T^3}{T_i^3}\,.
\end{align}

The Hubble rate is therefore given by
\begin{align}
    H_{EMD} = \sqrt{\frac{\rho_M}{3M_P^2}} = \sqrt{\frac{\rho_R^i}{3M_P^2}\frac{g_s(T)}{g_s(T_i)}\frac{T^3}{T_i^3}} = H_{RD}(T_i)\left(\frac{g_s(T)}{g_s(T_i)}\right)^{1/2}\left(\frac{T}{T_i}\right)^{3/2}\,.
\end{align}

As the EMD era is still isentropic, we still have $\rho_R\propto a^{-4}\propto g_e(T)T^4$.

\subsubsection*{For the EP era}

In the case of the late reheating, or the entropy production era, the situation
is more involved. Under the assumption that $\rho_M$ is still dominant and $\rho_M a^3\approx {\rm Const}$,
the Boltzmann equation in~\autoref{equ:BE_rho} can be solved semi-analytically:
\begin{align}
    \label{equ:rho_R_in_EP}
    \frac{d(\rho_R a^4)}{da} &\approx f\Gamma_M\frac{a^3\rho_M}{\sqrt{\frac{\rho_M}{3M_P^2}}} = \sqrt{3}M_Pf\Gamma_M\rho_M^{1/2}a^3 = \sqrt{3}M_Pf\Gamma_M\left(\rho_M^ia_i^3\right)^{1/2}a^{3/2} \nonumber \\
    \rho_R a^4 &\approx \frac{2}{5}\sqrt{3}M_Pf\Gamma_M(\rho_M^ia_i^3)^{1/2}a^{5/2} \nonumber \\
    \rho_R &\approx \frac{2}{5}\sqrt{3}M_Pf\Gamma_M\left(\rho_M^i\left(\frac{a_i}{a}\right)^3\right)^{1/2} \,.
\end{align}

Under the assumption that all decay products thermalize,
$\rho_R = \frac{\pi^2}{30}g_e(T)T^4$ and the Hubble parameter for EP era is given by
\begin{align}
    H_{EP} &= \sqrt{\frac{\rho_M}{3M_P^2}} = \sqrt{\frac{\rho_M^i a_i^3}{3M_P^2a^3}} \nonumber \\
           &= \frac{1}{\sqrt{3}M_P}\left(\rho_M^i\left(\frac{a_i}{a}\right)^3\right)^{1/2} \nonumber \\
           &= \frac{5}{2}\frac{1}{3M_P^2f\Gamma_M}\frac{\pi^2}{30}g_e(T)T^4 \,.
\end{align}

We have used Eq. (11) and the definition of $\rho_R$ in the last step of Eq. (12). The above formula can be further simplified by noting that the decay of the matter component is prominent when its decay width is comparable to the Hubble parameter, and that the decay happens at the end of EP\footnote{This assumption can be verified from~\autoref{fig:rho_s_T_H_evo}.}. In this case, we can write
$\Gamma_M = \kappa H_{RD}(T_r)$. We then reach for the Hubble parameter of EP era:
\begin{align}
    H_{EP}(T) = H_{RD}(T_r)\frac{5}{2}\frac{1}{f\kappa}\frac{g_e(T)}{g_e(T_r)}\left(\frac{T}{T_r}\right)^4\,.
\end{align}

Further, from~\autoref{equ:rho_R_in_EP} and the definition of $\rho_R\propto g_e(T)T^4$, we can see that the effect of the entropy production is to make the universe to cool down more slowly:
\begin{align}
\rho_R&\propto a^{-3/2}~\Rightarrow~T\propto g_e^{-1/4}(T)a^{-3/8}\,,
\end{align}
which defines a reheating period.

\subsubsection*{Continuity of the Hubble parameter}

The above results can be summarized as
\begin{subequations}
\begin{align}
    {\rm ERD} && [T_{RH},T_i] && H_{ERD} = \frac{\pi}{3}\sqrt{\frac{g_e(T)}{10}}\frac{T^2}{M_P} && \rho_R \,\propto\, a^{-4} \\
    {\rm EMD} && [T_i, T_e] && H_{EMD} = H_{RD}(T_i)\left(\frac{g_s(T)}{g_s(T_i)}\right)^{1/2}\left(\frac{T}{T_i}\right)^{3/2} && \rho_R\,\propto\, a^{-4} \\
    {\rm EP} && [T_e,T_r] && H_{EP} = H_{RD}(T_r)\frac{5}{2}\frac{1}{f\kappa}\frac{g_e(T)}{g_e(T_r)}\left(\frac{T}{T_r}\right)^4 && \rho_R \,\propto\, a^{-3/2} \\
    {\rm RD} && [T_r,\cdots] && H_{RD} = \frac{\pi}{3}\sqrt{\frac{g_e(T)}{10}}\frac{T^2}{M_P} && \rho_R\,\propto\, a^{-4} \,,
\end{align}
\end{subequations}
where $\cdots$ refers to temperatures below $T_r$ but still for a radiation-dominated universe.

The continuity of the Hubble parameter leads to the following constraints:
\begin{align}
    H_{ERD}(T_i) = H_{EMD}(T_i),\quad H_{EMD}(T_e)=H_{EP}(T_e),\quad H_{EP}(T_r) = H_{RD}(T_r) \,.
\end{align}

The first constraint is satisfied automatically in our procedure. The rest two lead to
\begin{subequations}
\begin{align}
    f\kappa &= \frac{5}{2}\\
    \left(\frac{T_e}{T_r}\right)^5 &= \frac{T_i}{T_r}\frac{g_s(T_e)}{g_s(T_i)}\frac{g_e(T_r)g_e(T_i)}{g_e^2(T_e)} \,,
\end{align}
\end{subequations}
where the second formula is used to obtain $T_e$ from $T_i$ and $T_r$.

Putting all these together, we finally have
\begin{subequations}
    \label{equ:Hubble_Splitting}
    \begin{align}
        {\rm ERD} && [T_{RH},T_i] && H_{ERD} = \frac{\pi}{3}\sqrt{\frac{g_e(T)}{10}}\frac{T^2}{M_P} && \rho_R\,\propto\, a^{-4} \\
        {\rm EMD} && [T_i, T_e] && H_{EMD} = H_{RD}(T_i)\left(\frac{g_s(T)}{g_s(T_i)}\right)^{1/2}\left(\frac{T}{T_i}\right)^{3/2} && \rho_R\,\propto\, a^{-4} \\
        {\rm EP} && [T_e,T_r] && H_{EP} = H_{RD}(T_r)\frac{g_e(T)}{g_e(T_r)}\left(\frac{T}{T_r}\right)^4 && \rho_R \,\propto\, a^{-3/2} \\
        {\rm RD} && [T_r,\cdots] && H_{RD} = \frac{\pi}{3}\sqrt{\frac{g_e(T)}{10}}\frac{T^2}{M_P} && \rho_R\,\propto\, a^{-4} \\
        {\rm With} && && \left(\frac{T_e}{T_r}\right)^5 = \frac{T_i}{T_r}\frac{g_s(T_e)}{g_s(T_i)}\frac{g_e(T_r)g_e(T_i)}{g_e^2(T_e)} \,.
    \end{align}
\end{subequations}

Note that there is no parameter directly related to the properties of the extra matter component in this parameterization. They are hidden behind the choice of $T_i$ and $T_r$, which determine respectively the beginning of the EMD era and the end of the EP era. The temperature $T_e$, which indicates the transition from the EMD to the EP era, is also determined by $T_i$ and $T_r$. Hence, in the study of cosmic relics considering an EMD era, $T_i$ and $T_r$ are chosen to be the free parameters.

\subsection{Dilution due to Entropy Production}

If the branching fraction of the extra matter component into the thermal bath is sufficient, it will inject energy/entropy into the thermal bath. Hence, as we have shown in the last section, the Hubble parameter behaves differently during the EP era. It is also shown in~\autoref{fig:rho_s_T_H_evo} that the total entropy increases during the EP era. It is interesting to have an estimation for the production of entropy.

From the above analysis, we have found
that $T\,\propto\, g_e^{-1/4} a^{-3/8}$ during the EP era. Then, for the total entropy we have
\begin{align}
    &S = sa^3 ~\propto~ g_s(T) T^3 a^3 = \frac{g_s(T)}{g_e^2(T)} T^3 T^{-8} = \frac{g_s(T)}{g_e^2(T)} T^{-5} \nonumber \\
\Rightarrow\quad & \frac{S_r}{S_e} = \left(\frac{T_e}{T_r}\right)^5 \frac{g_s(T_r)}{g_e^2(T_r)} \frac{g_e^2(T_e)}{g_s(T_e)} = \frac{T_i}{T_r}\frac{g_e(T_i)g_s(T_r)}{g_e(T_r)g_s(T_i)}\,.
\end{align}

Hence, the ratio between the entropy after and before the EP era is proportional to the ratio between $T_i$ and $T_r$. This ratio can be used to roughly estimate the dilution of the yield of cosmic relics. Physically, the more long-lived or feebly interacting the matter component, the longer the early matter-dominated era, and then the more entropy is produced. The amount of entropy produced is also related to the initial condition for the matter component \cite{Cosme:2020mck}, which can be therefore given in terms of $T_i$ and $T_r$.

\section{Boltzmann Equation}
\label{sec:BE}

The evolution of the number density for a given species of particle $k$ is governed by the following Boltzmann equation (BE):
\begin{align}\label{equ:BE}
a^{-3}\frac{d(n_ka^3)}{dt} &= \sum_{X\to Y}{r_k}\int \frac{d\Pi_X}{S_X} \int \frac{d\Pi_Y}{S_Y} (2\pi)^4\delta^{(4)}(p_X - p_Y)\nonumber \\
&\quad \hspace{1cm}\times \left(\prod_{i\in X} f_i\right)\left(\prod_{j\in Y}(1\pm f_j)\right)|\mathcal{M}(X\to Y)|^2 \,.
\end{align}

The summation runs over all processes involving particle $k$, which can be part of the multiparticle states $X$ and $Y$. Note that when expanding this summation, any reaction $X\to Y$ is in principle accompanied by its backreaction $Y\to X$. The factor $r_k = \bar n_k(-\bar n_k)$, for $k$ in $Y(X)$, counts the $\bar n_k>0$ particles $k$ in $Y(X)$. $\Pi_X$ and $\Pi_Y$ are the corresponding phase space factors,
\begin{align}
    d\Pi_X = \prod_{i\in X}\frac{d^4p_i}{(2\pi)^4}(2\pi\delta(p_i^2-m_i^2))\theta(p_i^0) =\prod_{i\in X}\frac{d^3p_i}{(2\pi)^32E_i} \,,
\end{align}
$S_X$ ($S_Y$) is the symmetry factor counting identical particles in $X$ ($Y$), $p_X = \sum_{i\in X} P_i$ ($p_Y = \sum_{i\in Y} P_i$) is the the sum of four-momenta in $X$ ($Y$), and $f_i$ are the distribution functions. Finally, $|\mathcal{M}(X\to Y)|^2$ is the squared matrix element for the process $X\to Y$ and is \textit{summed} over all internal degrees of freedom.

As usual, we consider three simplifying assumptions regarding the right-hand side of \autoref{equ:BE}. Firstly, we neglect the Pauli blocking/Bose enhancement factors ($(1\pm f_i)\to 1$). Secondly, we assume that the interacting species are kept in kinetic equilibrium at some temperature $T$, so that their out-of-equilibrium distribution functions are proportional to the equilibrium ones: $f_i = f_i^{eq}(T) \frac{n_i}{n_i^{eq}(T)}$ (see \cite{Binder:2021bmg} for a relaxation of this hypothesis). Finally, we assume Maxwell-Boltzmann statistics for the equilibrium distributions ($\prod_{i\in X}f_i^{eq}=e^{-\frac{E_X}{T}}$, with $E_X = \sum_{i\in X} E_i$). Under these assumptions, the BE simplifies to
\begin{align}
    a^{-3}\frac{d(n_ka^3)}{dt} &= \sum_{X\to Y} \mathcal{N}(k,X\to Y)\times \gamma(X\to Y)\,,
\end{align}
where we have defined the \textit{offset factor}
\begin{align}
\mathcal{N}(k,X\to Y) \equiv {r_k}\left(\prod_{i\in X} \frac{n_i}{n_i^{\rm eq}}\right)\,,
\end{align}
which quantifies the departure from chemical equilibrium, and the \textit{collision rate} (CR) density
\begin{align}
\gamma(X\to Y) \equiv \int \frac{d\Pi_X}{S_X} \int \frac{d\Pi_Y}{S_Y} (2\pi)^4\delta^{(4)}(p_X - p_Y)e^{-\frac{E_X}{T}}|\mathcal{M}(X\to Y)|^2\,,
\end{align}
which quantifies the number of interactions per unit of time and volume.

In the special case of CP-conserving interactions, the CR can be factorized and the offset
factor becomes an overall factor $\left(\prod_{i\in X}\frac{n_i}{n_i^{eq}}-\prod_{j\in Y}\frac{n_j}{n_j^{eq}}\right)$. However, the exact form of the offset depends on the problem one wants to study.

The left-hand side of BE can be further simplified as
\begin{align}
    \label{equ:BE_simplified}
    a^{-3}\frac{d(n_ka^3)}{dt} &= \frac{dn_k}{dt}+3Hn_k \nonumber \\
    &= s\frac{dY_k}{dt} + Y_k\frac{d s}{dt} + 3HY_k s \nonumber \\
    &= sH\left(-\frac{d\ln T}{d\ln a}\left(z\frac{dY_k}{dz}+Y_k \frac{d\ln s}{d\ln z}\right)+3Y_k\right) \nonumber \\
    &= \frac{sH}{\beta(T)}\left(z\frac{dY}{dz}+3(\beta(T) - g_s^*(T))Y_k\right)
\end{align}
where $Y_k=\frac{n_k}{s}$ is the yield of species $k$ and $z=\frac{M}{T}$ is a convenient time parameter, with $M$ some scale relevant for the problem at hand.
In the last equality, we have defined the dimensionless parameters as
\begin{subequations}
\begin{align}
\beta(T) &\equiv -\frac{d\ln a}{d\ln T}\\
g_s^*(T) &\equiv 1+ \frac{1}{3}\frac{d\ln g_s(T)}{d\ln T}
\end{align}
\end{subequations}

When treating the evolution in a `splitting' way with instantaneous transition as we did in~\autoref{sec:Hubble}, $\beta(T) = \frac{g_e^*(T)}{\beta_R}$ where $\beta_R$ is constant during each period and defined as
\begin{subequations}
\begin{align}
\rho_R &\propto g_e(T)T^4 \propto a^{-4\beta_R}\\
g_e^*(T) &\equiv 1+ \frac{1}{4}\frac{d\ln g_e(T)}{d\ln T}
\end{align}
\end{subequations}
For isentropic periods
(ERD/EMD/RD),  $\beta_R = 1$.
For the EP period, from~\autoref{equ:rho_R_in_EP}, $\beta_R = 3/8$. When treating the evolution according to the Boltzmann equation~\autoref{equ:evolution}, $\beta(T)$ can be obtained directly from the solution of $\rho_R$.

Finally, the Boltzmann fluid equation describing the evolution of a species $k$ in a universe that might have undergone an early matter-dominated era is given by
\begin{align}\boxed{
    z \frac{dY_k}{dz} + 3\left(\beta(T)-g_s^*(T)\right)Y_k = \frac{\beta(T)}{sH}\sum_{X\to Y} \mathcal{N}(k,X\to Y)\times \gamma(X\to Y)\,.}
\end{align}

The equation above is the primary equation of {\tt EvoEMD}. The summation runs over all processes involving particle $k$, and the offset and CR are calculated process by process. Moreover, a coupled set of such equations can be generated upon the requirement of the user. In what follows, we discuss the CR for n-body decays and $2\to n$ processes.

\subsection{Collision Rate for \texorpdfstring{$1\to n$}{2ton} Decay}

For $1\to n$ processes, we have
\begin{align}
    \gamma(1\to 23\cdots) &= \int\frac{d^3p_1}{(2\pi)^32E_1}e^{-E_1/T}\prod_j\int\frac{d^3p_j}{(2\pi)^32E_j}(2\pi)^4\delta^{(4)}(p_1-\sum_j p_j)\frac{|\mathcal{M}(1\to23\cdots)|^2}{S_Y} \nonumber \\
    &= \int \frac{p_1^2dp_1}{4\pi^2 E_1}e^{-E_1/T}\sumint|\mathcal{M}|^2 \nonumber \\
    &= \frac{m_1 T}{4\pi^2}K_1(m_1/T)\sumint|\mathcal{M}|^2\,,
\end{align}
where $K_1$ is the modified Bessel function of the second kind at 1st order. $\sumint|\mathcal{M}|^2$ represents the squared matrix element integrated over final state phase space, summed over all internal degrees of freedom and divided by all symmetry factors. For two body decays, we further have
\begin{align}
\sumint|\mathcal{M}|^2 &= \int d\Omega_{\rm CM}\frac{p_{23}}{16\pi^2 \sqrt{s}}\frac{|\mathcal{M}|^2}{S_{23}} \nonumber \\
&= (4\pi)\times\frac{\sqrt{\lambda(m_1^2,m_2^2,m_3^2)}/(2m_1)}{16\pi^2 m_1}\frac{|\mathcal{M}|^2}{S_{23}} \nonumber \\
&= \frac{1}{8\pi}\sqrt{\lambda\left(1,\frac{m_2^2}{m_1^2},\frac{m_3^2}{m_1^2}\right)}\frac{|\mathcal{M}|^2}{S_{23}} \,.
\end{align}
where $p_{ij}=\sqrt{\lambda(s,m_i^2,m_j^2)}/(2\sqrt{s})$ and $\lambda(x,y,z) = x^2 + y^2 + z^2 - 2xy - 2yz - 2zx$ is the K\"all\'en $\lambda$ function, $S_{23}$ is the symmetry factor for the final states.

\subsection{Collision Rate for \texorpdfstring{$2\to n$}{2ton} Scattering}
For $2\to n$ processes, we have
\begin{align}
    \gamma(12\to34\cdots) &= \int\frac{d^3p_1}{(2\pi)^32E_1}\frac{d^3p_2}{(2\pi)^32E_2}e^{-(E_1+E_2)/T} \nonumber \\
    &\quad \times\prod_j\int\frac{d^3p_j}{(2\pi)^32E_j}(2\pi)^4\delta^{(4)}\left(p_1+p_2-\sum_j p_j\right)\frac{|\mathcal{M}(12\to34\cdots)|^2}{S_XS_Y} \nonumber \\
    &= \int\frac{d^3p_1}{(2\pi)^32E_1}\frac{d^3p_2}{(2\pi)^32E_2}e^{-(E_1+E_2)/T} \sumint |\mathcal{M}|^2 \nonumber \\
    &= \frac{1}{8(2\pi)^4}\int ds dE_+ dE_- e^{-E_+/T} \sumint |\mathcal{M}|^2 \nonumber \\
    &= \frac{T}{2(2\pi)^4}\int d(\sqrt{s})s K_1\left(\frac{\sqrt{s}}{T}\right)\sqrt{\lambda\left(1,\frac{m_1^2}{s},\frac{m_2^2}{s}\right)}\sumint |\mathcal{M}|^2 \,,
\end{align}
where $E_\pm = E_1\pm E_2$ (see for instance \cite{Gondolo:1990dk} for details). For numerical convenience, we integrate over $\sqrt{s}>\max(m_X,m_Y)$ with $m_X (m_Y)$ being the sum over masses in state $X(Y)$.

Similar to the decay case, $K_1$ is the modified Bessel function of the second kind at 1st order. $\sumint |\mathcal{M}|^2$ is the squared matrix element integrated over final state phase space, summed over all internal degrees of freedom and divided by the symmetry factors.
For $2\to 2$ scattering process, we further have
\begin{align}
\sumint|\mathcal{M}|^2 &= \int d\Omega_{\rm CM}\frac{p_{34}}{16\pi^2 \sqrt{s}}\frac{|\mathcal{M}|^2}{S_{12}S_{23}} \nonumber \\
    &= \frac{\sqrt{\lambda(s,m_3^2,m_4^2)}/(2\sqrt{s})}{16\pi^2 \sqrt{s}}\int d\Omega_{\rm CM}\frac{|\mathcal{M}|^2}{S_{12}S_{23}} \nonumber \\
    &= \frac{1}{32\pi^2}\sqrt{\lambda\left(1,\frac{m_3^2}{s},\frac{m_4^2}{s}\right)}\int d\Omega_{\rm CM}\frac{|\mathcal{M}|^2}{S_{12}S_{23}} \,.
\end{align}

\section{Structure of {\tt EvoEMD}}
\label{sec:EvoEMD}

\begin{table}[]
\begin{tabular}{|r|l|}
\hline
\multicolumn{1}{|l|}{\textbf{src/EvoEMD/}}     & \textbf{Source files}
\\ \hline
ParameterBase.cpp                              & \begin{tabular}[c]{@{}l@{}}Defines functions of the {\tt Parameter\_Base} class, owned by the {\tt Parameter\_Factory} \\ (e.g.: {\tt DECLARE\_FREE\_PARAMETER}, {\tt RETRIEVE\_PARAMETER}, {\tt Set\_Value}, {\tt Get\_Value} \\  {\tt REGISTER\_PARAMETER}, {\tt Register\_Dependencies} )\end{tabular}

\\ \hline
ParticleBase.cpp                               &                    \begin{tabular}[c]{@{}l@{}}Defines functions of the {\tt Particle\_Base} class, owned by the {\tt Particle\_Factory} \\ (e.g.: {\tt REGISTER\_PARTICLE}, {\tt RETRIEVE\_PARTICLE}, {\tt REGISTER\_POI}, {\tt Get\_Mass})\end{tabular}

\\ \hline
ProcessBase.cpp                                &                      \begin{tabular}[c]{@{}l@{}}Defines functions of the {\tt Amplitude\_Base} class \\ (e.g.: {\tt REGISTER\_PROCESS}, {\tt Get\_Collision\_Rate}, {\tt Get\_Offset}, {\tt Update\_Value}, {\tt Update\_Amp}) \end{tabular}
\\
PhaseSpace.cpp                                 &                      \multicolumn{1}{c|}{\multirow{2}{*}{Define functions used in the calculation of the collision and offset factors}}

\\
CollisionRate.cpp                              &                    \multicolumn{1}{c|}{}                                                                                                                       \\ \hline
HubbleEvolution.cpp                            &                      Defines the functions used in the evaluation of the Hubble rate in the 'BE' mode

\\
HubbleSplitting.cpp                            &                      Defines the functions used in the evaluation of the Hubble rate in the 'Splitting' mode

\\ \hline
BoltzmannEquation.cpp                          &
\begin{tabular}[c]{@{}l@{}}
Defines functions used in the calculation of the set of Boltzmann equations for the POIs\\
(e.g.: {\tt BE}, {\tt Solve}, {\tt Get\_Yield\_at\_T\_End}, {\tt Get\_Omegah2\_at\_Today}, {\tt Dump\_Solution})
\end{tabular}

\\
RungeKutta.cpp                                 &                      \begin{tabular}[c]{@{}l@{}}
Implements the 4th order Runge-Kutta method with adaptive step size
\end{tabular}

\\ \hline
\multicolumn{1}{|l|}{\textbf{include/EvoEMD/}} & \textbf{Header files}                                                                      \\ \hline
EvoEMD.h                                       &
Includes all necessary components one needs to use EvoEMD; should be called in the main files

\\ \hline
ParameterBase.h                                &
Declares classes, functions, and variables used in the corresponding source file
\\
...                                            &
\\ \hline
HubbleBase.h                                   &                      Declares the {\tt Hubble\_Base} and {\tt Hubble\_Factory} classes
\\ \hline
\end{tabular}
\caption{The EvoEMD library. Users do not need to change this part of the code for studying the evolution of relics in the context of an early matter-dominated universe. All functions and parameters defined in the EvoEMD library can be called and retrieved in the main files of the Model folder.}
\label{tab:EvoEMDlib}
\end{table}

With the {\tt EvoEMD} framework, users can study the evolution of relics throughout a universe which may have undergone an early matter-dominated era. The structure of the {\tt EvoEMD} library is shown in \autoref{tab:EvoEMDlib}. There are mainly five important parts of {\tt EvoEMD}:
\begin{itemize}
    \item {\tt Parameters}: Including all free and derived parameters used in the calculation.
    \item {\tt Particles}: Including all particles involved in the calculation. Some of them can be claimed {\it particle-of-interest} (POI) which will enter the Boltzmann equation.
    \item {\tt Processes}: Including all processes involved in the Boltzmann equation.
    \item {\tt Hubble}: Calculating the Hubble parameter at given temperature.
    \item {\tt Boltzmann Equation}: Building the Boltzmann equation automatically according to the information user implemented, and solving the Boltzmann equation using the 4th order Runge-Kutta method with adaptive step size.
\end{itemize}

In what follows, we describe each of these parts and discuss how to use them.

\subsection{Parameters}
\label{sec:parameters}

The parameters include free and derived parameters. The {\tt EvoEMD} framework assumes all parameters to be real. All parameters declared will be hold and owned by a factory object (\cppin{Parameter_Factory}), such that the users do not need to worry about the scope of the object. At the end of the program, the \cppin{Parameter_Factory} will take care the resources used by all parameters.

\subsubsection*{Free Parameters}

The free parameters can be declared and retrieved according to their names. Hence, users should assign a unique name for each parameter. The framework will just prompt a warning if there is a duplicated name, but will continue running with unpredictable results.
We provide the macro {\cppin{DECLARE_FREE_PARAMETER(param_name, value)}} to declare free parameters, where \cppin{param_name} is the user-specified parameter name and \cppin{value} is the initial value. To retrieve the pointer (\cppin{Parameter_Base*}, which is the base class for all parameters) to the parameter, one can use the macro {\cppin{Parameter_Base *ptr = RETRIEVE_PARAMETER(param_name)}}. With this \cppin{Parameter_Base} pointer, one can set and obtain the value of corresponding parameters:
\begin{cpp}
ptr->Set_Value(new_value);
REAL value = ptr->Get_Value();
\end{cpp}

\subsubsection*{Derived Parameters}

In general, besides free parameters, we will have derived parameters. Users need to define such parameter through a class derived from \cppin{Parameter_Base}. Users are required to define the default constructor, and \cppin{void Update_Value(REAL input)} member function. However, the functionality can be extended by defining any other member functions.

When defining the constructor, one has to define the name for the parameter and also indicate the free parameters it depends on. In \cppin{void Update_Value(REAL input)},
users are free to do anything to compute the value of the parameter, and store the result to variable \cppin{value}. An example of how to register derived parameters is given in the file {\cppin{Parameters.cpp}} of the leptogenesis model described in \autoref{sec:Appendix_LG}, which defines the parameter class \cppin{param_GammaN1}. The definition of a parameter class should not be confused with the definition of the parameter itself. To define/declare the parameter itself, one needs to register the derived parameter in the model header file using the class name:
\begin{cpp}
REGISTER_PARAMETER(param_GammaN1); // REGISTER_PARAMETER(className)
Parameter_Base *ptr = RETRIEVE_PARAMETER(GammaN1) // Obtain the parameter using its name (GammaN1)
\end{cpp}

Notice that, in the constructor of the derived parameter, we register the dependencies. This tells the program that current parameter depends on those parameters. Then, whenever we update the value of the dependencies, the current parameter will automatically recalculate its value only when we want to use it.

\subsection{Particles}
\label{sec:particles}
All particles in the system are objects of their own classes inherited from an abstract class \cppin{Particle_Base}, and will be hold and owned by a factory class \cppin{Particle_Factory}. A particle has the following properties:

\begin{itemize}
\item Name: \cppin{string}, the name of the particle.
\item PID: \cppin{int}, unique id for the particle. The PID is used to distinguish different particles.
\item DOF: \cppin{int}, internal degree of freedom of the particle. The dof of particle and anti-particle are counted separately.
\item Mass: \cppin{Parameter_Base} pointer to the mass parameter, for massless particle it is \cppin{nullptr}.
\item Width: \cppin{Parameter_Base} pointer to the width parameter, for stable particle (or particle we don't care the width) it is \cppin{nullptr}.
\item Pseudo: \cppin{bool}, a flag indicating whether it is a real particle or not.
\end{itemize}

In general, we want to use the Boltzmann equation to track the evolution of the number density of some particles. However, in some special case, we may track the difference of number densities. Hence, in the {\tt EvoEMD} framework, we add a flag (\cppin{Pseudo}) to indicate whether it is a real particle or a net quantum number for instance, such as the difference of particle and antiparticle. There is an important difference between \cppin{Pseudo=true} and \cppin{Pseudo=false} that for a real particle, the number density cannot be negative, while for a pseudo-particle, it can. Note that we are ignorant about whether the particle is self-conjugate. For a self-conjugate particle, there is no ambiguity. However, for particle that is not self-conjugate, we assume that if users want to track the evolution of the total number density of both particle and anti-particle, they should provide the total collision rate through the {\tt Process} discussed below including both particle and anti-particle. If, in any case, users just want to track the number density of particle (or anti-particle) alone, only the corresponding processes should be implemented.

There are two kinds of particles, fermion and boson. The main difference is the equilibrium number density, where a fermion follows the Fermi-Dirac distribution and a boson follows the Bose-Einstein distribution. For simplicity, in the {\tt EvoEMD} framework, we only use the corresponding distributions for the massless case. For massive particles, we always use Maxwell distribution. Hence, one should define fermions and bosons separately:
\begin{cpp}
// REGISTER_PARTICLE(Type, name, pid, dof, mass_ptr, width_ptr, pseudo);
// mass_ptr and width_ptr have default value as nullptr; pseudo has default value as false
REGISTER_PARTICLE(Fermion, N1, 900001, 2, RETRIEVE_PARAMETER(MN1), RETRIEVE_PARAMETER(GammaN1), false);
REGISTER_PARTICLE(Boson, S, 900025, 2, RETRIEVE_PARAMETER(MS), nullptr, false);
REGISTER_PARTICLE(Fermion, dL, 900011, 2 * 2, nullptr, nullptr, true); // Lepton number
\end{cpp}

Some of the particles can be claimed as particle-of-interest
(POI) using its PID which will enter the Boltzmann equation, after one registers the particle into \cppin{Particle_Factory}:

\begin{cpp}
// REGISTER_POI(PID, INIT_THERMAL_STATUS);
REGISTER_POI(900001, false);
REGISTER_POI(900025, true);
\end{cpp}
where the second argument indicates whether the particle is initially thermalized with the SM bath.
This flag can be changed at any time before one starts solving the Boltzmann equation.

All particles registered into the system can be retrieved from the system by using its PID:
\begin{cpp}
auto *ptr = RETRIEVE_PARTICLE(900001);
\end{cpp}

With the pointer to a particle, we can obtain/set its properties:
\begin{cpp}
int PID = ptr->Get_PID();
int dof = ptr->Get_DOF();
string name = ptr->Get_Name();
REAL mass = ptr->Get_Mass();
bool massless = ptr->Is_Massless();
bool pesudo = ptr->Is_Pseudo();
bool thermal = ptr->Get_Init_Thermal_Status();

REAL T = 100;
REAL neq = ptr->Get_Equilibrium_Number_Density_at_T(T);
REAL Yeq = ptr->Get_Equilibrium_Yield_at_T(T);

ptr->Set_Init_Thermal_Status(false);
ptr->Set_Mass(200);
\end{cpp}

\subsection{Processes}
\label{sec:processes}
In order to build up the Boltzmann equation, one should provide the CR and offset for all the relevant processes. The {\tt EvoEMD} framework will automatically calculate the CR from $\sumint |\mathcal{M}|^2$ using the formula provided in~\autoref{sec:BE} for decay and scattering processes. Hence, {\tt EvoEMD} provides an abstract \cppin{Amplitude_Base} class for the user to provide the $\sumint |\mathcal{M}|^2$ and offset. Any process should be implemented as derived class of \cppin{Amplitude_Base} where three member functions should be overridden:
\begin{cpp}
void Update_Value(REAL input);
void Update_Amp(REAL sqrt_shat);
REAL Get_Offset(REAL T, int PID);
\end{cpp}

The function \cppin{Update_Value} is inherited from \cppin{Parameter_Base} which is used to update any $\sqrt{s}$-independent variable. Note that \cppin{Amplitude_Base} is inherited from \cppin{Parameter_Base} and can be treated as derived parameter. Hence, as long as one register the dependencies for the amplitude, any $\sqrt{s}$-independent variable will be updated if the value of any of its dependencies changes. The function \cppin{Update_Amp} is used to calculate the value for $\sumint|\mathcal{M}|^2$ at given $\sqrt{s}$, and store the result into variable \cppin{amp_res}. The last function \cppin{Get_Offset} is used to calculate the offset in the Boltzmann equation related to current process. It will only be a function of the temperature. However, for the Boltzmann equation of different species, we will also have different offset for the same process. Hence, the second argument \cppin{PID} is used to indicate the species. With the implemented amplitude, another class \cppin{Process} is then used to further calculate the CR according to the formulae in~\autoref{sec:BE} using the {\tt Cuba} library~\cite{Hahn:2004fe,Hahn:2014fua}. Users do not need to know details about \cppin{Process} but just need to register the process in the model header file using the macro {\cppin{REGISTER_PROCESS(amp_class_name)}}, where \cppin{amp_class_name} is the class name for the process inherited from \cppin{Amplitude_Base}.

\subsection{Hubble Parameter}

The Hubble parameter is calculated either according to~\autoref{equ:Hubble_Splitting} (`Splitting') or by solving the Boltzmann equation~\autoref{equ:evolution} (`BE'). In either method, the calculation depends on two input parameters $T_i$ and $T_r$. In the `BE' method, the branch fraction $f$ is also needed. In the `Splitting' method, $T_i$ and $T_r$ are respectively the temperature of the starting point of the EMD era and the end point of the EP era. In the `BE' method, they are defined as the temperatures when $\rho_M=\rho_R$. The definition of $T_i$ can be matched between these two methods, while there may be a small shift in $T_r$.

Both methods are stored in the object of a factory class \cppin{Hubble_Factory}. An extra parameter \cppin{HubbleMethod} is used to control which method is to be used. By default, the `Splitting' and `BE' methods are provided with \cppin{HubbleMethod = 0} and \cppin{HubbleMethod = 1} respectively. Relevant parameters have already been declared in the {\tt EvoEMD} framework:
\begin{cpp}
DECLARE_FREE_PARAMETER(Ti, 1e14);
DECLARE_FREE_PARAMETER(Tr, 10);
DECLARE_FREE_PARAMETER(BR, 1.0); // `f' in Eq. (2b)
DECLARE_FREE_PARAMETER(HubbleMethod, 0);
\end{cpp}

One can change the value of these parameters at any time, the Hubble parameter calculation will be automatically updated accordingly. The Hubble parameter calculator can be accessed at any time using
\begin{cpp}
Hubble_Base *hc = Hubble_Factory::Get_Hubble_Calculator(id);
REAL hubble = hc->Get_Hubble_at_T(temperature);
REAL dlna_dlnT = hc->Get_dlna_dlnT_at_T(temperature);
\end{cpp}
where \cppin{id >= 0} is an optional argument and will override (but not replace) the value from \cppin{HubbleMethod}. When omitting this argument, the method corresponding to \cppin{HubbleMethod} will be used. Whenever \cppin{id} or \cppin{HubbleMethod} is invalid, `Splitting' method will be used. Further, two member functions are provided by \cppin{Hubble_Base}:
\begin{itemize}
    \item \cppin{REAL Get_Hubble_at_T(const REAL T)}: obtain the Hubble parameter at given temperature \cppin{T}.
    \item \cppin{REAL Get_dlna_dlnT_at_T(const REAL T)}: ontain the $\frac{d\ln a}{d\ln T}$ ($-\beta(T)$) at given temperature \cppin{T}.
\end{itemize}

Users are free to implement their own calculators for the Hubble parameter. In this case, they need to define their own calculator class inherited from \cppin{Hubble_Base}, and overload the two member functions \cppin{REAL Get_Hubble_at_T(const REAL T)} and \cppin{REAL Get_dlna_dlnT_at_T(const REAL T)}. We also provide an extra class \cppin{Hubble_Evolution} which can be used to solve the Boltzmann equation~\autoref{equ:evolution} with arbitrary initial conditions. In fact, `BE' method is implemented based on \cppin{Hubble_Evolution} with the initial conditions given by~\autoref{equ:rhoM_rhoR_initial}. Users are free to provide any initial conditions through
\begin{cpp}
Hubble_Evolution::Solve(REAL x_ini, REAL Y1_ini, REAL Y2_ini, REAL BR = 1);
\end{cpp}

In this sense, {\tt EvoEMD} can also be used to study the evolution of cosmic relics in a broader range of cosmological models.

\subsection{Boltzmann Equation}

The {\tt EvoEMD} framework builds up the Boltzmann equation automatically, according to the POI and the processes (including CR and offset) associated with the corresponding POI as

\begin{align}
z \frac{dY_k}{dz} + 3\left(\beta(T)-g_s^*(T)\right)Y_k = \frac{\beta(T)}{sH}\sum_{X\to Y} \mathcal{N}(k,X\to Y)\times \gamma(X\to Y)\,.
\end{align}

The program solves the Boltzmann equation in terms of $z=\frac{M}{T}$, where $M$ is some important scale in the problem. This scale should be provided as a \cppin{Parameter_Base} pointer which should be an already declared parameter. The Boltzmann equation can be therefore declared as a function of this scale:

\begin{cpp}
Parameter_Base *scale = RETRIEVE_PARAMETER(MN1); // Get the scale
BoltzmannEquation BE(scale); // Define the Boltzmann equation with scale
REAL ss = scale->Get_Value();

REAL T_BEGIN = ss*100;
REAL T_END = ss/100;
BE.Set_T_Range(T_BEGIN,T_END); // Set the range of the temperature to solve the BE

BE.Solve(step_size,tolerance); // Solve the Boltzmann equation
\end{cpp}

When setting the range of the temperature, we will automatically set the initial conditions for the yield of the POIs according to their initial thermal status. The Boltzmann equation then can be solved by calling \cppin{BE.Solve(step_size,tolerance)} where \cppin{step_size} is the initial step size in solving the Boltzmann equation, and \cppin{tolerance} is the largest relative error allowed in each step when solving the Boltzmann equation.

Currently, we implement the 4th order Runge-Kutta method with adaptive step size. In the cosmological applications, we tailored the step control method such that when the collision happens too fast, the program will try to trace the equilibrium number density for real particle. With this improvement, users do not need to check the thermalization condition. The thermalization will be automatically achieved whenever it is possible. Further, a cache system will be built for each involved process, storing temporarily the CR result to accelerate the computations. At the end of each call of \cppin{Solve}, the cache will be cleaned to avoid conflicts among the scans with different parameters.

Finally, users can obtain the yield ($Y = \frac{n}{s}$) at the end of the evolution by
\begin{cpp}
VD yend = BE.Get_Yield_at_T_End();
\end{cpp}

We also provide a function to obtain the current relic density $\Omega^0_i h^2$ of the POIs:
\begin{cpp}
VD omegah2 = BE.Get_Omegah2_at_Today();
\end{cpp}

The $\Omega^0_i h^2$ is calculated in the following way:
\begin{align}
    \Omega_i^0 = \frac{\rho_i^0}{\rho^c} = \frac{m_i n_i^0}{\rho^c}\,,
\end{align}
where $\rho^c = 8.098h^2\times 10^{-11}\,{\rm eV}^4$ is the critical density and $\rho^0_i$ is the energy density of the POI particle $i$ today, which must be non-relativistic. After freeze-out/freeze-in and after the entire EP period, the POI number density simply falls off as $a^{-3}$. Since the entropy density also scales as $a^{-3}$ for temperatures below $T_r$, the yield itself is unchanged. Then the energy density is
\begin{align}
    n_i^0 = Y_i^{0} s_{0} = Y_i^{\rm end}s_0\,,
\end{align}
where $s_{0}\simeq 2.22\times 10^{-38}$ GeV$^3$ is the entropy density today.
Finally, we have
\begin{align}
    \Omega_i^0 h^2 = \frac{m_i Y_i^{\rm end}s_0}{8.098\times10^{-11}{\rm eV}^4} =\frac{m_i}{\rm GeV}\frac{Y_i^{\rm end}}{3.643\times10^{-9}}.
\end{align}

\section{Usage and Examples}
\label{sec:Examples}

\subsection{Installation and Instructions}

To install {\tt EvoEMD}, users first need to have \cppin{GSL}~\cite{GSL} as well as \cppin{Cuba}~\cite{Hahn:2004fe,Hahn:2014fua} in their system. \cppin{GSL} comes with most Unix-like systems or can be easily installed through package manage systems. \cppin{Cuba} should be previously compiled. The corresponding directory of \cppin{Cuba} should be provided when building up {\tt EvoEMD}. With these dependencies satisfied, one can install {\tt EvoEMD} with
\begin{bash}
git clone https://github.com/ycwu1030/EvoEMD.git
cd EvoEMD
mkdir build; cd build
cmake -DCMAKE_INSTALL_PREFIX=INSTALL_PATH -DCUBA_ROOT=CUBA_PATH ../
make
make install # Optional
\end{bash}
where \cppin{INSTALL_PATH} is the directory where to install the package, \cppin{CUBA_PATH} is the {\tt Cuba} directory. The option for {\tt Cuba} directory is mandatory, while the install prefix is optional. \cppin{make install} is optional which installs the headers, library, and some other components into \cppin{INSTALL_PATH}. However, even without \cppin{make install}, users can still link their own program to {\tt EvoEMD}, as the library is in \cppin{build/lib} and all headers are in \cppin{SOURCE_DIR/include}.

To use {\tt EvoEMD}, users are required to build their own `model' by providing the parameters (free and derived), particles (as well as POI), and the amplitudes for all relevant processes. The definition of free parameter and particles (POIs) is straightforward as illustrated in~\autoref{sec:parameters} and~\autoref{sec:particles} respectively. However, one should define the classes inherited from \cppin{Parameter_Base} for derived parameters and from \cppin{Amplitude_Base} for process amplitude. Then, users can define all physical objects including parameters, particles, processes, and solve the Boltzmann equation for the system by linking to the {\tt EvoEMD}.

\begin{figure}[!tb]
    \centering
    \includegraphics[scale=0.35]{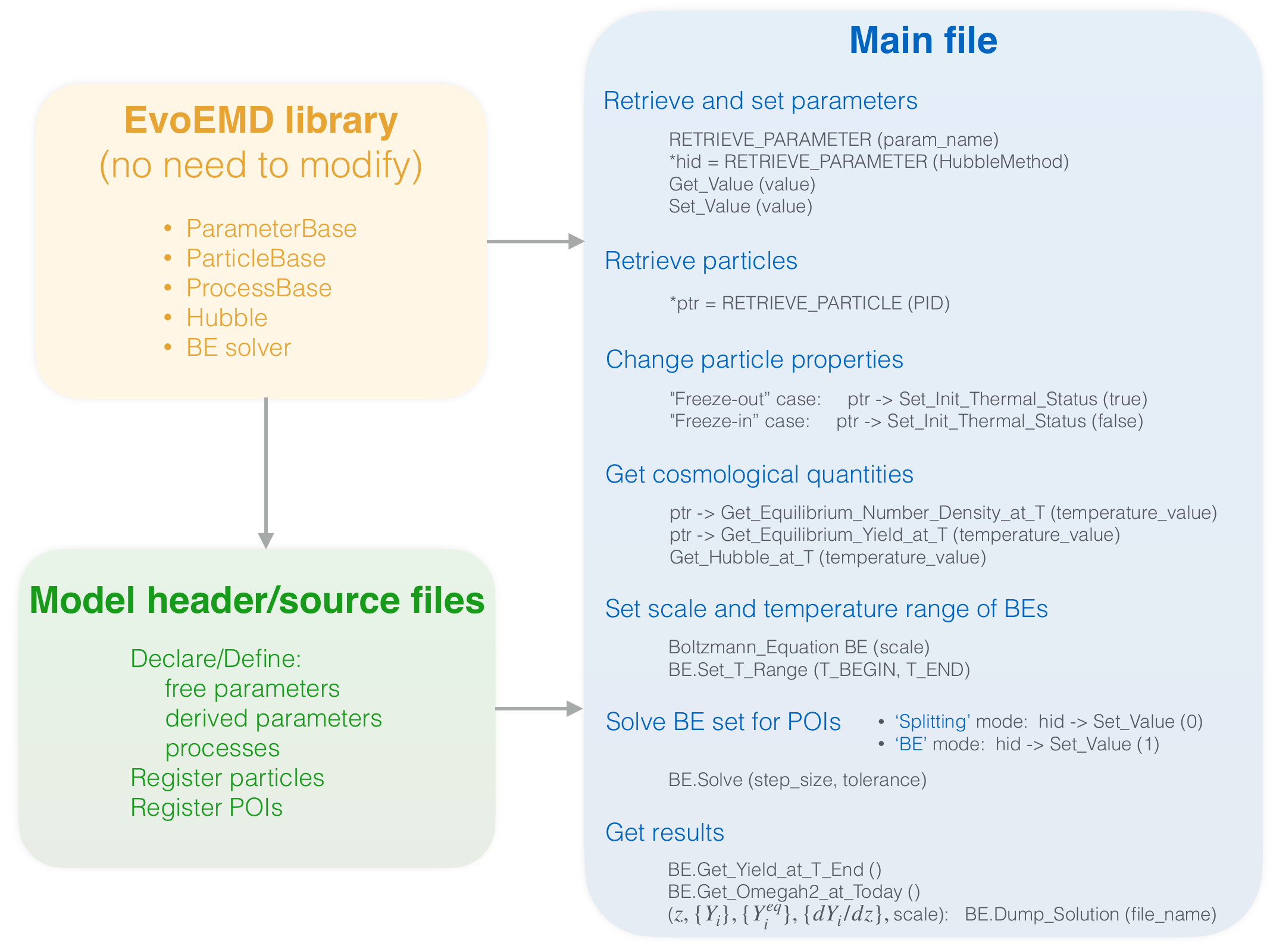}
    \caption{Schematic overview of the {\tt EvoEMD} structure and usage. In the model main file, users can call functions defined in the {\tt EvoEMD} library and in the model source files and change all relevant parameters.}
    \label{fig:EvoEMD_flowchart}
\end{figure}

In \autoref{fig:EvoEMD_flowchart}, we show an schematic overview of the usage of {\tt EvoEMD}. Users do not need to modify the {\tt EvoEMD} library, located in {\cppin{SOURCE_DIR/src}} and {\cppin{SOURCE_DIR/include}}. In \cppin{SOURCE_DIR/Models}, new models can be included and linked to {\tt EvoEMD} by editing the {\tt CMakeList} files provided in \cppin{SOURCE_DIR/} and \cppin{SOURCE_DIR/Models/ToyDM/} for instance. Model header, source, and main files are stored \cppin{SOURCE_DIR/Models/MODEL_FOLDER}. In the main file, users can retrieve the parameters defined in the source model files and change their values as desired. In particular, the duration of the EMD era ($T_i, T_r$) can be changed. The method for computing the Hubble parameter can also be changed with the pointer \cppin{*hid}. By retrieving particles, it is possible to change their initial thermalization status and to compute cosmological quantities such as their equilibrium number density. The scale $M$ and the temperature range for the Boltzmann equations must be set in the main file. The Boltzmann equations for the set of POIs are then solved with a given step size and tolerance chosen directly in the main file.

After running the code, the solution for the yields of the POIs are dumped into output files located in {\cppin{SOURCE_DIR/build/bin}}. The first column of an output file is $z = M/T$, the following columns are the yields of the POIs at $T$, the equilibrium yields of the POIs, the differential yields $dY/dz$, and the last column is the scale at which the equations were solved. The yields appear in the order that they were declared. Users can also run scans over the parameter space using the resulting quantities such as the final yield {\cppin{Get_Yield_at_T_End}} and the final relic density {\cppin{Get_Omegah2_at_Today()}}, and store results in their own files.

In the following, we describe two toy models (provided in \cppin{SOURCE_DIR/Models}) to show the capabilities of {\tt EvoEMD}. The first example is a toy dark matter model, which provides both freeze-out and freeze-in dark matter calculations. The second example is a toy leptogenesis model, which starts with the heavy right hand neutrino either thermalized or non-thermalized.

\subsection{Toy Dark Matter Model}

Let us consider a simple dark matter model consisting of a real scalar dark matter candidate $\chi$ with mass $m_\chi$ \cite{Silveira:1985rk,McDonald:1993ex,Burgess:2000yq,Cline:2013gha}. It communicates with the SM fields with a coupling strength $\lambda$ to the SM Higgs doublet $\Phi$:
\begin{align}
    \mathcal{L}\supset \frac{\lambda}{2} \chi^2 |\Phi|^2.
\end{align}

For the purpose of showing the capabilities of {\tt EvoEMD}, only one process (as well as its T-conjugated one) is involved in our toy model: $\chi\chi\leftrightarrow \Phi \Phi^\dagger$. The Boltzmann equation for $\chi$ is then
\begin{align}\label{Eq:BE_ToyDM}
    z \frac{dY_\chi}{dz} + 3\left(\beta(T)-g_s^*(T)\right)Y_\chi &= \frac{\beta(T)}{sH}\sum_{X\to Y} \mathcal{N}(\chi,X\to Y)\times \gamma(X\to Y) \nonumber \\
    &= \frac{\beta(T)}{sH}\left[2\left(1 - \frac{Y_\chi^2}{(Y_\chi^{\rm eq})^2}\right)\right]\gamma(\chi\chi\to \Phi \Phi^\dagger)\,.
\end{align}

In the second equality, we assumed that CP is conserved and summed over $\chi\chi\to\Phi \Phi^\dagger$ and $\Phi \Phi^\dagger\to \chi\chi$.

From the Lagrangian, we obtain the squared matrix element summed over internal d.o.f (also divided by symmetry factor) as
\begin{align}
    \sum |\mathcal{M}|^2 &= \frac{1}{2}\times 2\times \lambda^2 = \lambda^2\nonumber \\
\Rightarrow\quad \sumint |\mathcal{M}|^2 &= \frac{\lambda^2}{8\pi}\sqrt{\lambda\left(1,\frac{m_\Phi^2}{s},\frac{m_\Phi^2}{s}\right)} \,.
\end{align}

This model is implemented into the following files, located in {\cppin{SOURCE_DIR/Models/ToyDM}}:
\begin{cpp}
Amplitudes.h
Amplitudes.cpp
ToyDM.h
ToyDM.cpp
\end{cpp}

First, in \cppin{Amplitudes.h} and \cppin{Amplitudes.cpp}, we define the amplitude class for the process $\chi\chi\leftrightarrow \Phi \Phi^\dagger$, inheriting from \cppin{Amplitude_Base}. In \cppin{Amplitudes.cpp}, one provides all the information needed to compute the right-hand side of \autoref{Eq:BE_ToyDM} (the integrated squared amplitude entering the reaction rate density and the offset factor). In the header file {\tt ToyDM.h}, all physical parameters, particles, and processes objects in the model are declared/registered. In particular, the DM particle is registered as POI. Finally, we can track the evolution of the yield/number density of the DM ($\chi$) in the main source file \cppin{ToyDM.cpp}. In \autoref{sec:Appendix_DM}, we provide detailed code excerpts of these files.

The example has already been compiled and linked to {\tt EvoEMD}\footnote{See the {\tt CMakeList} files provided in \cppin{SOURCE_DIR/} and \cppin{SOURCE_DIR/Models/ToyDM/}.} when one build up {\tt EvoEMD}. One can run this example by
\begin{bash}
EvoEMD/build$ make # Optional, re-make if one makes any change in the example
EvoEMD/build/bin$ ./ToyDM
\end{bash}

\begin{figure}
    \centering
    \includegraphics[width=.9\textwidth]{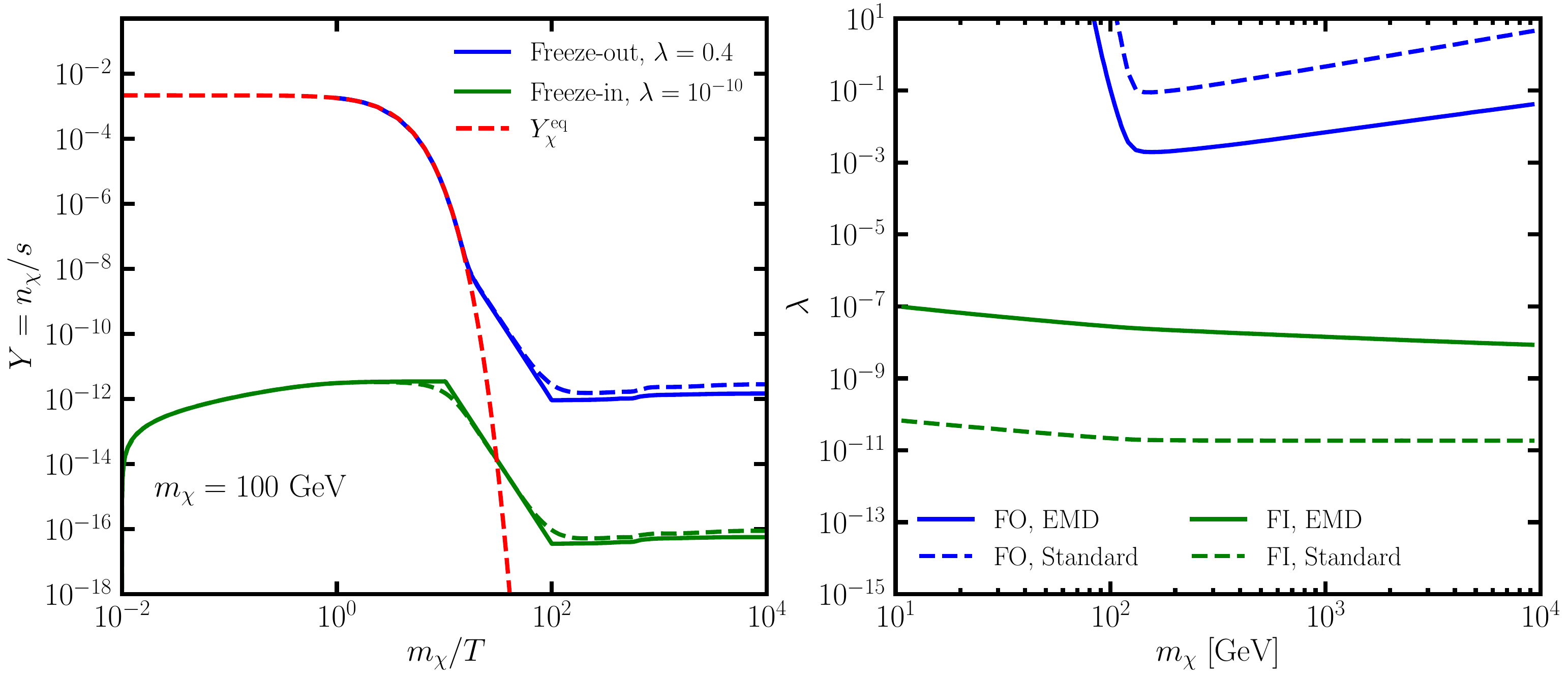}
    \caption{Left: The evolution of the yield for the DM $\chi$, with $m_\chi = 100$ GeV. The red dashed line indicates the equilibrium yield. The blue and green solid lines are respectively for the freeze-out case ($\lambda = 0.4$) and the freeze-in case ($\lambda = 10^{-10}$) using the `Splitting' method for Hubble parameter. The corresponding dashed lines are for the cases where the `BE' method for Hubble parameter is used. An EMD era was set to start at $T_i=10^5$ GeV and finish at $T_r=1$ GeV, with a visible dilution of the yields due to the EP era. Right: The contours of observed relic density in the $m_\chi$-$\lambda$ plane using the `Splitting' method. Dashed contours are for the standard cosmology, while solid contours are for the Universe with an EMD era with $T_i=10^5$ GeV and $T_r=1$ GeV.}
    \label{fig:ToyDM}
\end{figure}

In the left panel of~\autoref{fig:ToyDM}, we plot the solution for the yield of the DM candidate, $Y_\chi$, as function of $z = m_\chi/T$ when $m_\chi = 100$ GeV, $T_i=10^5$ GeV and $T_r=1$ GeV. We consider the evolution of $Y_\chi$ in both the freeze-out case (blue curves), in which $\lambda = 0.4$ and $\chi$ is initially thermalized with $\Phi$, and freeze-in case (green curves), in which $\lambda = 10^{-10}$ and $\chi$ is initially absent and thermalization is never achieved. Note that the freeze-out evolution starts following the yield at equilibrium $Y_\chi^{eq}$ (dashed red curve), which can be obtained within {\tt EvoEMD} by calling the function \cppin{Get_Equilibrium_Yield_at_T(T)} and is given in the output file if one calls the function {\cppin{Dump_Solution(file_name)}}. In turn, one must ensure that the freeze-in finishes while the $Y_\chi \ll Y^{eq}_\chi$. The solid curves are for the case where the `Splitting' method is used for the calculation of the Hubble parameter, while the corresponding dashed curves are for the case where the `BE' method is used. The kinks in both freeze-out and freeze-in cases come from the dilution due to the entropy production. The transition between different periods is smooth when the `BE' method is used. However, the `Splitting' method also provides good estimations and spends much less time. Hence, in default, {\tt EvoEMD} will use the `Splitting' method.

In the right panel of~\autoref{fig:ToyDM}, we show contours of correct relic density of $\chi$ in the parameter plane $m_\chi-\lambda$. This scan was performed in the main file with the function \cppin{Get_Omegah2_at_Today()}, for the case of freeze-out (blue contours) and freeze-in (green contours). Dashed curves are for the standard cosmological scenario of a RD universe. Solid contours are for an EMD era starting at $T_i = 10^5$GeV followed by an EP period finishing at $T_r = 1$ GeV. As we can see, an EMD allows for the WIMPs to be more weakly interacting, possibly evading the current bounds, while avoiding overproduction. In the case of the FIMPs, an EMD era requires larger couplings, facilitating their phenomenology.

The freeze-out contours in~\autoref{fig:ToyDM} only takes into account the quartic coupling annihilation $\chi \chi \to \Phi^\dagger \Phi$. As shown, this process becomes available for $m_\chi > m_\Phi$. In the freeze-in case, Higgs doublets in the thermal bath can always produce pairs of $\chi$, but when $m_\chi \gg m_\Phi$, the cross-section (and then the relic contours) becomes mostly independent on $m_\chi$. In realistic models of singlet scalar dark matter, the trilinear coupling $\chi \chi h$, with $h$ the Higgs boson, becomes possible after the electroweak symmetry breaking. Thus, s-channel processes must be considered in the calculation of the relic density, significantly changing the contours of correct relic density in both cases of freeze-out \cite{Burgess:2000yq,Cline:2013gha} and freeze-in \cite{Chu:2011be,Blennow:2013jba}.

\subsection{Toy Leptogenesis Model}

Now, let us consider a simple leptogenesis model in which a net lepton number is generated through the CP-violating out-of-equilibrium decay of a heavy right-handed neutrino $N$. The main interaction is
\begin{align}
\mathcal{L} &\supset -\left(\lambda e^{i\alpha} \overline{L_L}\tilde{\Phi}N_R + h.c.\right) - \frac{1}{2}m_N \bar{N}N \,.
\end{align}

Note that, in this toy model, we simply assume that the phase $\alpha$ in the coupling cannot be rotated away by field redefinition. For simplicity, $L$ and $\Phi$ are all assumed to be massless, while $N_R$ is a Majorana fermion with mass $m_N\sim 10^{13}$ GeV. The relevant processes include the $\Delta L = 1$ process $N\to L\Phi$ and the $\Delta L = 2$ processes $L\Phi \to L^c \Phi^\dagger$ and $LL\to\Phi^\dagger \Phi^\dagger$. The Boltzmann equations that control the evolution of $Y_N$ and $Y_L$ are
\begin{subequations}
\begin{align}
    z\frac{dY_N}{dz} + 3\left(\beta(T)-g_s^*(T)\right)Y_N &= \frac{\beta(T)}{sH}\left[\left(1-\frac{Y_N}{Y_N^{\rm eq}}\right)\gamma^{N}_{L\Phi} - \frac{Y_L}{2Y_\ell^{\rm eq}}\delta\gamma^{N}_{L\Phi}\right], \\
    z\frac{dY_L}{dz} + 3\left(\beta(T)-g_s^*(T)\right)Y_L &= \frac{\beta(T)}{sH}\left[-\left(1-\frac{Y_N}{Y_N^{\rm eq}}\right)\delta\gamma^N_{L\Phi} - \frac{Y_L}{2Y_\ell^{\rm eq}}\left(\gamma^N_{L\Phi}+ 2\gamma^{\prime L\Phi}_{L^c\Phi^\dagger}+4\gamma^{LL}_{\Phi^\dagger\Phi^\dagger}\right)\right] \nonumber \\
    & \approx \frac{\beta(T)}{sH}\left[-\left(1-\frac{Y_N}{Y_N^{\rm eq}}\right)\delta\gamma^N_{L\Phi} - \frac{Y_L}{2Y_\ell^{\rm eq}}\gamma^N_{L\Phi}\right]\,,
\end{align}
\end{subequations}
where $Y_L$ is the yield for total lepton number (lepton minus anti lepton), while $Y_\ell^{\rm eq}$ is the yield at equilibrium for lepton. Further, we introduce the collision rate of CP-conserving and CP-violating:
\begin{align}
    \gamma^X_Y &\equiv \gamma(X\to Y) + \gamma(\bar X\to \bar Y),\nonumber \\
    \delta \gamma^X_Y &\equiv \gamma(X\to Y) - \gamma(\bar X\to \bar Y).
\end{align}

For $2\to2$ process with s-channel propagators, $\gamma'$ indicates the subtracted collision rate where the real intermediate state contribution is subtracted to avoid double counting. In this toy example, we further simplify the discussion by ignoring the $2\to2$ wash-out terms as they are one order in $\lambda$ higher than that of the decay process. In a realistic analysis, one should seriously analyze the corresponding influence of these wash-out effects.

For the CP-conserving case, ignoring the loop corrections, we have
\begin{align}
(\text{CP-conserving})\quad    \sumint |\mathcal{M}|^2 &= \frac{1}{8\pi}\times\sqrt{\lambda(1,0,0)}|\mathcal{M}|^2 = \frac{1}{8\pi}\times \left(4\lambda^2m_N^2\right).
\end{align}

CP-violation is raised only at loop level with the interference between the tree level amplitude and the loop-induced amplitude. To further simplify the implementation in this example, we introduce the lepton asymmetry
\begin{align}
\epsilon \equiv \frac{\Gamma(N\to L\Phi)-\Gamma(N\to L^c\Phi^\dagger)}{\Gamma(N\to L\Phi)+\Gamma(N\to L^c\Phi^\dagger)}
\end{align}
as a free parameter. In general, it is a function of $\lambda$, $\alpha$ and $m_N$. Then, we have for the CP-violating part
\begin{align}
(\text{CP-violating})\quad    \sumint|\mathcal{M}|^2 = \frac{1}{8\pi}\times\left(4\epsilon\lambda^2m_N^2\right).
\end{align}

This toy example is implemented into the following files, located in {\cppin{SOURCE_DIR/Models/ToyLeptogenesis}}:
\begin{cpp}
Parameters.h
Parameters.cpp
Amplitudes.h
Amplitudes.cpp
ToyLG.h
ToyLG.cpp
\end{cpp}

In \cppin{Parameters.h} and \cppin{Parameters.cpp}, we define a class, inherited from \cppin{Parameter_Base}, for the decay width of $N$, the only derived parameter in the example~\footnote{Note that in current example, we didn't use the width. However, in a realistic study, when the $2\to2$ processes are considered, the width will be used for the propagators.}.

Then, in \cppin{Amplitudes.h} and \cppin{Amplitudes.cpp}, the CP-conserving and CP-violating amplitudes for $N\to L\Phi$ are implemented (see detailed code excerpts in \autoref{sec:Appendix_LG}).

All the physical objects are declared in \cppin{ToyLG.h}. In this case, one must register both the particle $N$ and the lepton number $L$ (pseudo particle) as POIs. Finally, we can track the evolution of the yield/number density of the heavy neutrino and lepton number in the main file \cppin{ToyLG.cpp}.

In the main file, users can retrieve and set different values for the free parameters, which automatically updates the value of the derived parameters (in this example, the width of $N$). Since in this case both $N$ and $L$ were registered as POIs, the solver will automatically compute the coupled Boltzmann equations for $Y_N$ and $Y_{L}$. After setting the scale for the evolution parameter, in this case the mass of the heavy neutrino, one sets the temperature range and the initial status of the POIs. The default initial status is thermalized.

\begin{figure}[!tbp]
\centering
\includegraphics[width=0.6\textwidth]{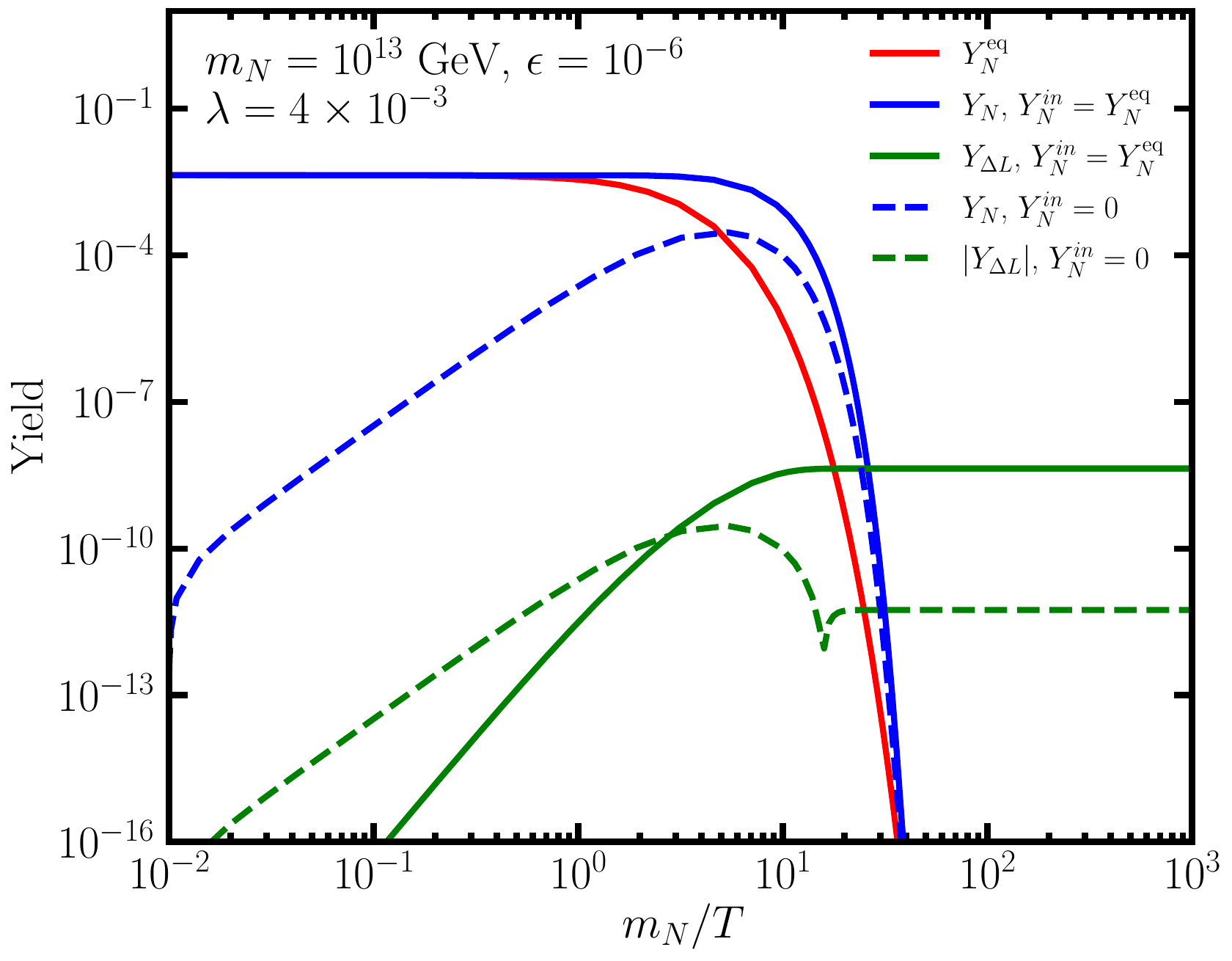}
\caption{The evolution of the yields for heavy neutrino (blue lines) and lepton number (green lines). The red line indicates the equilibrium yield of the heavy neutrino. Two cases are shown: Solid lines are for the case where $N$ starts thermalized $Y_N^{in}=Y_N^{\rm eq}$, while the dashed lines are for the case where $N$ starts out-of-equilibrium $Y_N^{in}=0$. For $Y_N^{in}=0$, the yield for the lepton number is negative and changes its sign around $z=\frac{m_N}{T}\approx 16$.}
\label{fig:ToyLG}
\end{figure}

In ~\autoref{fig:ToyLG}, we show the evolution of heavy neutrino and lepton number in two cases: $N$ initially thermalized, $Y_N^{in}=Y_N^{\rm eq}$ (solid blue and green), and not thermalized, $Y_N^{in}=0$ (dashed blue and green). The solutions are respectively dumped in the output files \cppin{ToyLG_FO_Result.txt} and \cppin{ToyLG_FI_Result.txt} once we run the main file provided.

The example has already been compiled and linked to {\tt EvoEMD} when one build up {\tt EvoEMD}. One can run this example by
\begin{bash}
EvoEMD/build$ make # Optional, re-make if one makes any change in the example
EvoEMD/build/bin$ ./ToyLG
\end{bash}

\section{Summary}
\label{sec:Summary}

The universe might have undergone an early matter-dominated (EMD) era prior to the big bang nucleosynthesis. This is a direct consequence of many well-motivated BSM scenarios in which decoupled feebly interacting fields dominate the cosmic expansion before their complete decay. This possibility significantly impacts the establishment of cosmic relics, such as dark matter and the matter-antimatter asymmetry. Moreover, the phenomenology of such relics also changes significantly. In this work, we introduce {\tt EvoEMD}, a new public tool to evaluate, to the best of our knowledge for the first time, the evolution of cosmic relics in a universe with or without an EMD era.

An EMD era brings the need to track number densities of cosmic relics together with the energy densities of the extra matter and the radiation components. In {\tt EvoEMD}, we provide a simplified calculation
by suitably splitting the Hubble rate according to its temperature-dependence (`Splitting'). This allows us to focus on the Boltzmann equations for number densities, with general collision terms, while the effect of the EMD era is incorporated into the Hubble rate. The reheating period which might follow the EMD era is also taken into account. We also provide a method to directly solve the coupled Boltzmann equations for the energy densities (`BE').
{\tt EvoEMD} can be therefore extended in order to account for the establishment of cosmic relics during the inflationary reheating period. We have also provided general expressions for collision terms, without the usual assumption of CP-conservation.

{\tt EvoEMD} is written in {\tt C++} and has dedicated frameworks for the particle physics model implementation and for the evaluation of the cosmological quantities. Numerical integration of collision terms are performed with the {\tt Cuba} library. Two examples are provided to illustrate the usage/ability of {\tt EvoEMD}. In the toy DM model, a scalar DM interacts with the Higgs bilinear via quartic coupling. Users can evaluate how the DM number density evolves if DM is initially thermalized with the Higgs bosons (freeze-out case) and when DM is initially absent in the early universe and never attains thermalization (freeze-in case). In the toy DM model, we also checked the difference between `Splitting' and `BE' methods. It is found that `Splitting' method provides a quite good approximation while spends much less time. We also provided a toy leptogenesis model in which the CP-violating decay of a heavy sterile neutrino generates a net lepton number further translated into a matter-antimatter asymmetry. In this example, we have shown how the coupled Boltzmann equations for the yields of the heavy neutrino and the lepton number are solved in {\tt EvoEMD}, in both cases of heavy neutrinos initially thermalized or never thermalized.

Arbitrary particle physics models can be implemented by specifying the relevant particles, parameters, and the squared matrix elements for relevant processes. Then {\tt EvoEMD} can be used to track the relics of any particles of interest such that the impact of an EMD era in the early Universe can be evaluated. Another independent package will be released to simplify the implementation of arbitrary models in the future.

\section*{Acknowledgments}

M.D. acknowledges the support of the Natural Sciences and Engineering Research Council
of Canada. Y.W. thanks the U.S.~Department of Energy for the financial support, under grant number DE-SC 0016013.

\appendix

\section{Model files for the dark matter model}
\label{sec:Appendix_DM}

In this appendix, we provide a detailed description of the implementation of the dark matter model, provided in \cppin{SOURCE_DIR/Models/ToyDM/}.

All information regarding the processes relevant for the dark matter evolution is written in the files {\cppin{Amplitudes.h}} and {\cppin{Amplitudes.cpp}}. They are implemented as follows

\begin{cpp}
// ==================================== Amplitudes.h ====================================
#ifndef _TOY_DM_PROCESSES_H_
#define _TOY_DM_PROCESSES_H_
#include "EvoEMD/EvoEMD.h" // This header includes all necessary components one needs to use EvoEMD

// All amplitude class should be derived from Amplitude_Base
class XX_HH_Amp : public EvoEMD::Amplitude_Base {
private:
    REAL Sub1; // Template s-independent variable

public:
    XX_HH_Amp();
    ~XX_HH_Amp(){};

    virtual void Update_Value(REAL input) override;
    virtual void Update_Amp(REAL sqrt_shat) override;
    virtual REAL Get_Offset(REAL T, int PID) override;
};
#endif  //_TOY_DM_PROCESSES_H_
\end{cpp}

\begin{cpp}
// ==================================== Amplitudes.cpp ====================================
#include "Amplitudes.h"
#include <cmath>
#include "EvoEMD/EvoEMD.h"

using namespace EvoEMD; // All components in EvoEMD are declared under namespace EvoEMD

// Initialize the base class with the name
XX_HH_Amp::XX_HH_Amp() : Amplitude_Base("XX_HH") {
    // Obtain the particles involved in this process by using their user-defined PID
    // Amplitude.h and Amplitude.cpp define the class, but not the object. So users can use any particle or parameter they want here, but when they actually define the object of this class, the corresponding particles and parameters must have been already defined
    Particle_Base *p_dm = RETRIEVE_PARTICLE(900001);
    Particle_Base *p_H = RETRIEVE_PARTICLE(900025);

// Define the initial and final states of the process
    INITIAL.push_back(p_dm);
    INITIAL.push_back(p_dm);
    FINAL.push_back(p_H);
    FINAL.push_back(p_H);
    N_INITIAL = INITIAL.size();
    N_FINAL = FINAL.size();

// Amplitude_Base is inherited from Parameter_Base, through which we can handle the parameter dependencies. One can register any dependencies here
    auto *ptr_lam = RETRIEVE_PARAMETER(Lam);
    Register_Dependencies(ptr_lam);
}

// In this function, one update the value of any s-independent variables
void XX_HH_Amp::Update_Value(REAL input) {
    REAL lam = RETRIEVE_PARAMETER(Lam)->Get_Value();
    Sub1 = pow(lam, 2);
}

// This function updates the amplitude and stores the results in amp_res;
void XX_HH_Amp::Update_Amp(REAL sqrt_shat) {
    REAL mH = RETRIEVE_PARTICLE(900025)->Get_Mass();
    REAL MH2 = mH * mH;
    REAL s = sqrt_shat * sqrt_shat;
    REAL kallen_sqrt = sqrt(Kallen_Lam(1.0, MH2 / s, MH2 / s));
    amp_res = Sub1 * kallen_sqrt / 8.0 / M_PI;
}

REAL XX_HH_Amp::Get_Offset(REAL T, int PID) {
    // The first argument is the temperature, the second argument is the PID
    // As the coefficient for BE depends on which particle we are considering
    if (PID == 900001) {
        // Usually, the coefficient depends on the current yield of the particle;
        // So we provide a member data in each particle to store current yield;
        Particle_Base *pp = RETRIEVE_PARTICLE(900001);
        REAL Y = pp->Yield;
        REAL YeqT = pp->Get_Equilibrium_Yield_at_T(T);

        // Note that, for massive particles, when T is much lower than the mass, the Yeq may underflow,
        // Hence, we directly set it to zero. One needs to check such situation.
        if (YeqT == 0) return 0;
        REAL res = 2.0 * (1.0 - pow(Y / YeqT, 2));

        // Further, we also store 1-Y/Yeq for each particle
        // When Y is equal to Yeq, due to numerical issues, when calculating the offset, it will not return zero
        // Hence, for Y close to Yeq, we recommend the use of Delta_Yield_Ratio instead.
        if (fabs(res) < 1e-5) {
            res = 2.0 * (2 * pp->Delta_Yield_Ratio);
        }

        return res;
    } else {
        return 0;
    }
}
\end{cpp}

The header file \cppin{ToyDM.h} is implemented as follows
\begin{cpp}
// ==================================== ToyDM.h ====================================
#ifndef _TOY_DM_H_
#define _TOY_DM_H_

#include "Amplitudes.h"
#include "EvoEMD/EvoEMD.h"
using namespace EvoEMD;

// Free Parameters
DECLARE_FREE_PARAMETER(Lam, 0.4);
DECLARE_FREE_PARAMETER(MX, 100);
DECLARE_FREE_PARAMETER(MH, 125);

// All Particles
REGISTER_PARTICLE(Boson, X, 900001, 1, RETRIEVE_PARAMETER(MX), nullptr);
REGISTER_PARTICLE(Boson, H, 900025, 2, RETRIEVE_PARAMETER(MH), nullptr);

// Particles entering the Boltzmann Equation
// We first assume it is thermalized at the beginning, it can be changed later
REGISTER_POI(900001, true);

// Register Processes
// Note that processes must be registered after involved particles and parameters
REGISTER_PROCESS(XX_HH_Amp);

#endif  //_TOY_DM_H_
\end{cpp}

\begin{figure}[!ht]
    \centering
    \includegraphics[width=\textwidth]{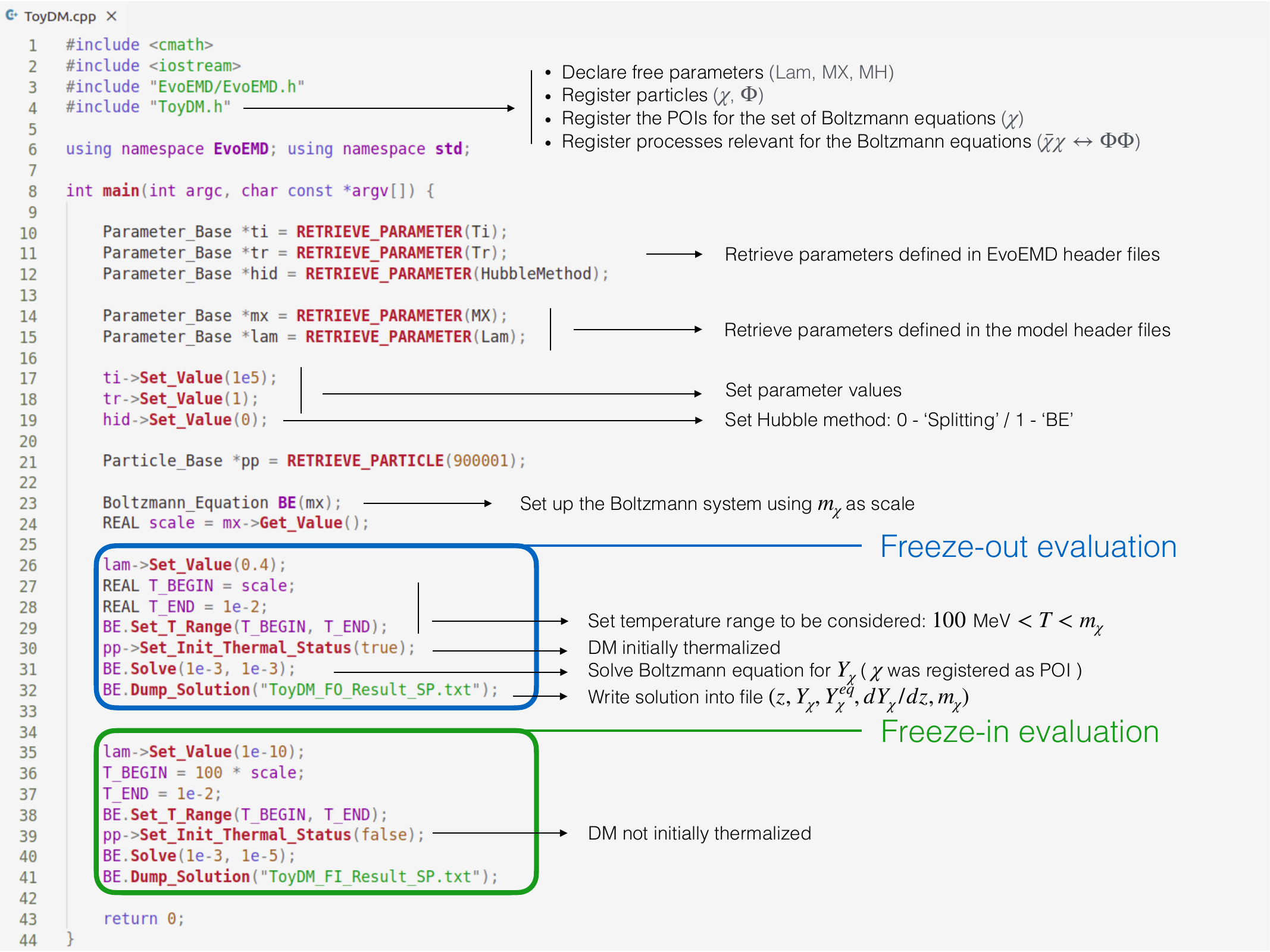}
    \caption{The main file for the dark matter model implemented in {\tt EvoEMD} (located in EvoEMD/Models/ToyDM).}
    \label{fig:ToyDM_print}
\end{figure}

In \autoref{fig:ToyDM_print}, we describe each part of the main file \cppin{ToyDM.cpp}. In a few code lines, users can evaluate the freeze-out and the freeze-in of a dark matter candidate in the context of an EMD era. Users can retrieve and set different values for all relevant parameters before running the code. Then, they should specify the temperature range of the set of Boltzmann equations (in this particular case, the Boltzmann equation for the evolution of $Y_\chi$) and define whether the POIs are initially thermalized. After that, the solver for the Boltzmann equations is called with {\cppin{BE.Solve(step_size,tolerance)}}. The function {\tt Dump\_Solution}, whose argument is the name of an output file, is then called to dump the solution of the Boltzmann equations. After running the code provided, the solution for the yields are dumped into four output files located in {\cppin{SOURCE_DIR/build/bin}}:
\begin{cpp}
ToyDM_FO_Result_SP.txt
ToyDM_FI_Result_SP.txt
ToyDM_FO_Result_BE.txt
ToyDM_FI_Result_BE.txt
\end{cpp}
from which one can find the evolution of the yield of the dark matter $\chi$. The files \cppin{ToyDM_FO_Result_BE.txt} and \cppin{ToyDM_FI_Result_BE.txt} correspond to the case where the `BE' method is chosen (not shown in \autoref{fig:ToyDM_print}).

\section{Model files for the leptogenesis model}
\label{sec:Appendix_LG}

Let us now describe the implementation of the leptogenesis model, which is provided in \cppin{SOURCE_DIR/Models/ToyLeptogenesis/}.

First, we define the class for derived parameters, in this case the decay width of $N$, which is inherited from \cppin{Parameter_Base} in \cppin{Parameters.h} and \cppin{Parameters.cpp}
\begin{cpp}
// ==================================== Parameters.h ====================================
#ifndef _TOY_LEPTOGENESIS_PARAMETER_H_
#define _TOY_LEPTOGENESIS_PARAMETER_H_

#include "EvoEMD/EvoEMD.h"
using namespace EvoEMD;

class param_GammaN1 : public Parameter_Base {
public:
    param_GammaN1();

    virtual void Update_Value(REAL input) override; // For derived parameter, one must override this function
};
#endif  //_TOY_LEPTOGENESIS_PARAMETER_H_
\end{cpp}

\begin{cpp}
// ==================================== Parameters.cpp ====================================
#include "Parameters.h"

// We assign the name ``GammaN1'' to this parameter which will be used to retrieve it
param_GammaN1::param_GammaN1() : Parameter_Base("GammaN1") {
    // GammaN1 depends on the mass MN1 and coupling Lam
    Parameter_Base* p_mn1 = RETRIEVE_PARAMETER(MN1);
    Parameter_Base* p_lam = RETRIEVE_PARAMETER(Lam);

    Register_Dependencies(p_mn1, p_lam); // Register the dependencies
}

void param_GammaN1::Update_Value(REAL input) {
// The input can be ignored for derived parameters
// Get the value of the two free parameters
    REAL mn1 = RETRIEVE_PARAMETER(MN1)->Get_Value();
    REAL lam = RETRIEVE_PARAMETER(Lam)->Get_Value();
// In this function, one updates the value for the parameter and store it in `value':
    value = lam * lam * mn1 / 8 / M_PI;
}
\end{cpp}

All information regarding the processes relevant for leptogenesis is written in the files {\cppin{Amplitudes.h}} and {\cppin{Amplitudes.cpp}}. They are implemented as follows

\begin{cpp}
// ==================================== Amplitudes.h ====================================
#ifndef _TOY_LEPTOGENESIS_AMP_H_
#define _TOY_LEPTOGENESIS_AMP_H_
#include "EvoEMD/EvoEMD.h"

class N_LPhi_Amp_CPC : public EvoEMD::Amplitude_Base {
private:
    REAL Sub1;
public:
    N_LPhi_Amp_CPC();
    virtual void Update_Value(REAL input) override;
    virtual void Update_Amp(REAL sqrt_shat) override;
    virtual REAL Get_Offset(REAL T, int PID) override;
};

class N_LPhi_Amp_CPV : public EvoEMD::Amplitude_Base {
private:
    REAL Sub1;
public:
    N_LPhi_Amp_CPV();
    virtual void Update_Value(REAL input) override;
    virtual void Update_Amp(REAL sqrt_shat) override;
    virtual REAL Get_Offset(REAL T, int PID) override;
};
#endif  //_TOY_LEPTOGENESIS_AMP_H_
\end{cpp}
\begin{cpp}
// ==================================== Amplitudes.cpp ====================================
#include "Amplitudes.h"
using namespace EvoEMD;
N_LPhi_Amp_CPC::N_LPhi_Amp_CPC() : Amplitude_Base("NLPhi_CPC") {
    Particle_Base *p_N1 = RETRIEVE_PARTICLE(900001);
    Particle_Base *p_l = RETRIEVE_PARTICLE(900011);
    Particle_Base *p_phi = RETRIEVE_PARTICLE(25);

    FINAL.push_back(p_l);
    FINAL.push_back(p_phi);
    INITIAL.push_back(p_N1);

    N_INITIAL = INITIAL.size();
    N_FINAL = FINAL.size();

    Parameter_Base *ptr_lam = RETRIEVE_PARAMETER(Lam);
    Parameter_Base *ptr_mn1 = RETRIEVE_PARAMETER(MN1);
    Register_Dependencies(ptr_lam, ptr_mn1);
}
void N_LPhi_Amp_CPC::Update_Value(REAL input) {
    REAL lam = RETRIEVE_PARAMETER(Lam)->Get_Value();
    REAL mn1 = RETRIEVE_PARAMETER(MN1)->Get_Value();

    Sub1 = 4 * lam * lam * mn1 * mn1;
}

void N_LPhi_Amp_CPC::Update_Amp(REAL sqrt_shat) {
    // Final states l and phi are massless, so sqrt(lam(1,0,0)) = 1;
    amp_res = 1.0 / 8.0 / M_PI * Sub1;
}

REAL N_LPhi_Amp_CPC::Get_Offset(REAL T, int PID) {
    // The first argument is the temperature, the second argument is the PID;
    // As the coefficient for BE depends on which particle we are considering;
    if (PID == 900001) {
        Particle_Base *pp = RETRIEVE_PARTICLE(900001);
        REAL Y = pp->Yield;
        REAL YeqT = pp->Get_Equilibrium_Yield_at_T(T);
        if (YeqT == 0) return 0;
        REAL res = (1.0 - Y / YeqT);
        if (fabs(res) < 1e-5) {
            res = pp->Delta_Yield_Ratio;
        }
        return res;
    } else if (PID == 900011) {
        Particle_Base *pp = RETRIEVE_PARTICLE(900011);
        REAL Y = pp->Yield;
        REAL YeqT = pp->Get_Equilibrium_Yield_at_T(T);
        if (YeqT == 0) return 0;
        REAL res = -Y / YeqT / 2.0;
        return res;
    } else {
        return 0;
    }
}

N_LPhi_Amp_CPV::N_LPhi_Amp_CPV() : Amplitude_Base("NLPhi_CPV") {
    Particle_Base *p_N1 = RETRIEVE_PARTICLE(900001);
    Particle_Base *p_l = RETRIEVE_PARTICLE(900011);
    Particle_Base *p_phi = RETRIEVE_PARTICLE(25);

    FINAL.push_back(p_l);
    FINAL.push_back(p_phi);
    INITIAL.push_back(p_N1);

    N_INITIAL = INITIAL.size();
    N_FINAL = FINAL.size();

    auto *ptr_lam = RETRIEVE_PARAMETER(Lam);
    auto *ptr_mn1 = RETRIEVE_PARAMETER(MN1);
    auto *ptr_eps = RETRIEVE_PARAMETER(Eps);
    Register_Dependencies(ptr_lam, ptr_mn1, ptr_eps);
}

void N_LPhi_Amp_CPV::Update_Value(REAL input) {
    REAL lam = RETRIEVE_PARAMETER(Lam)->Get_Value();
    REAL mn1 = RETRIEVE_PARAMETER(MN1)->Get_Value();
    REAL eps = RETRIEVE_PARAMETER(Eps)->Get_Value();

    Sub1 = 4 * eps * lam * lam * mn1 * mn1;
}

void N_LPhi_Amp_CPV::Update_Amp(REAL sqrt_shat) {
    amp_res = 1.0 / 8.0 / M_PI * Sub1;
}

REAL N_LPhi_Amp_CPV::Get_Offset(REAL T, int PID) {
    REAL res = 0;
    if (PID == 900001) {
        Particle_Base *pp = RETRIEVE_PARTICLE(900011);
        REAL Y = pp->Yield;
        REAL YeqT = pp->Get_Equilibrium_Yield_at_T(T);
        if (YeqT == 0) return 0;
        res = -Y / YeqT / 2.0;
    } else if (PID == 900011) {
        Particle_Base *pp = RETRIEVE_PARTICLE(900001);
        REAL Y = pp->Yield;
        REAL YeqT = pp->Get_Equilibrium_Yield_at_T(T);
        if (YeqT == 0) return 0;
        res = -(1.0 - Y / YeqT);
        if (fabs(res) < 1e-5) {
            res = -pp->Delta_Yield_Ratio;
        }
    }
    return res;
}
\end{cpp}

Now, we define all the physical objects in \cppin{ToyLG.h}
\begin{cpp}
// ==================================== ToyLG.h ====================================
#ifndef _TOY_LEPTOGENESIS_H_
#define _TOY_LEPTOGENESIS_H_
#include "Amplitudes.h"
#include "EvoEMD/EvoEMD.h"
#include "Parameters.h"
using namespace EvoEMD;

// Declare free Parameters
DECLARE_FREE_PARAMETER(MN1, 1e13);
DECLARE_FREE_PARAMETER(Lam, 4e-3);
DECLARE_FREE_PARAMETER(Eps, 1e-6);

// Register derived parameters
REGISTER_PARAMETER(param_GammaN1);

// Register particles (real or pseudo)
REGISTER_PARTICLE(Fermion, N1, 900001, 2, RETRIEVE_PARAMETER(MN1), RETRIEVE_PARAMETER(GammaN1));
REGISTER_PARTICLE(Fermion, dL, 900011, 2 * 2, nullptr, nullptr, true);
REGISTER_PARTICLE(Boson, Phi, 25, 2, nullptr, nullptr);

// Register particles entering the Boltzmann Equations
REGISTER_POI(900001, 1);
REGISTER_POI(900011, 0);

// Register Processes relevant for the Boltzmann Equations
REGISTER_PROCESS(N_LPhi_Amp_CPC);
REGISTER_PROCESS(N_LPhi_Amp_CPV);

#endif  //_TOY_LEPTOGENESIS_H_
\end{cpp}

\begin{figure}[!tb]
    \centering
    \includegraphics[width=\textwidth]{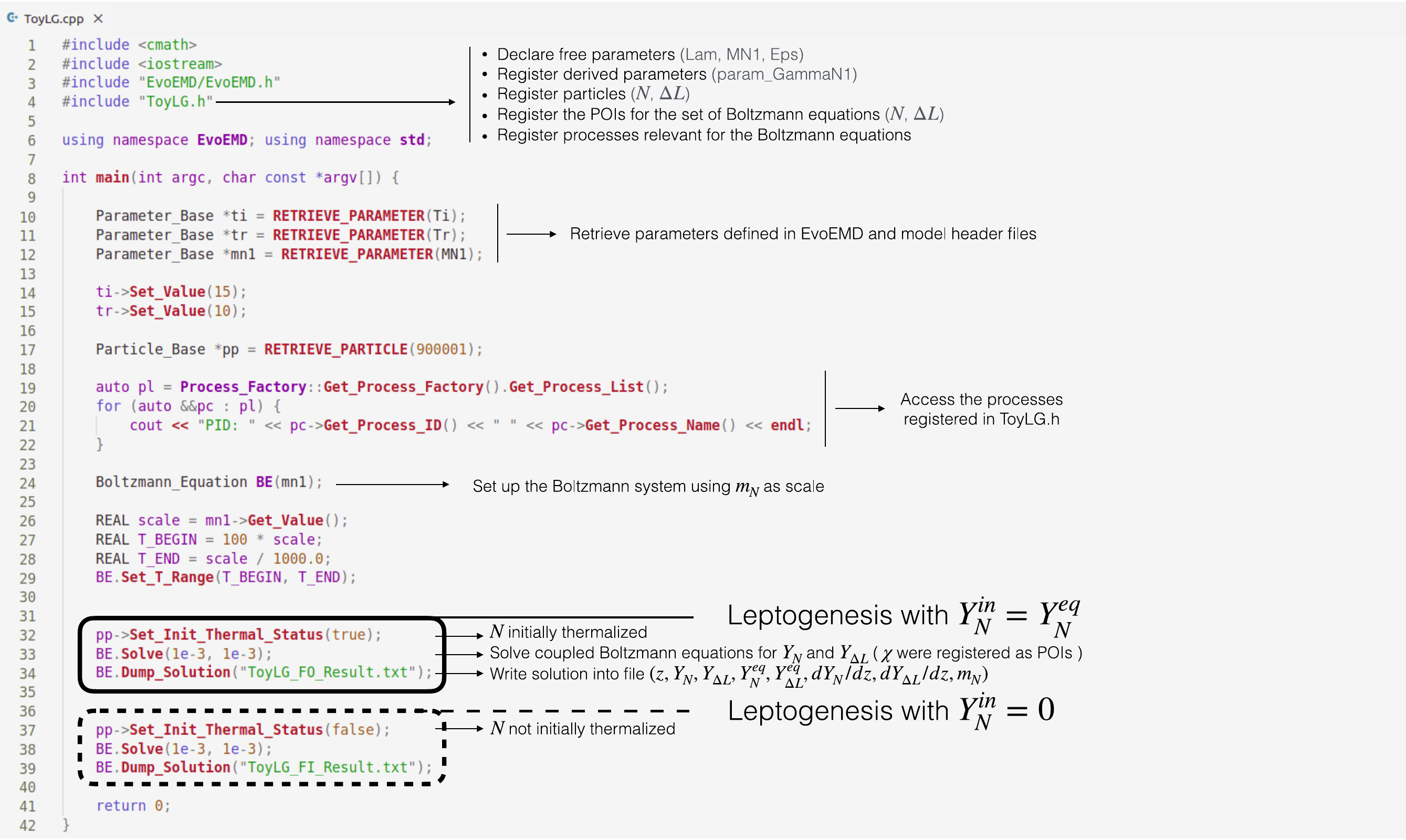}
    \caption{The main file for the leptogenesis model implemented in {\tt EvoEMD} (located in EvoEMD/Models/ToyLeptogenesis).}
    \label{fig:ToyLG_print}
\end{figure}

In \autoref{fig:ToyLG_print}, we show the main file for the leptogenesis model. In a few code lines, users can evaluate the coupled evolution of a heavy neutrino and the net lepton number generated by its decay. After running this code, the solution for the yields are dumped into files located in {\cppin{SOURCE_DIR/build/bin}}.

\end{document}

%% file: macros.tex
\usepackage{listings}
\newcommand*\cppin{\lstinline[language=c++]}

\lstnewenvironment{cpp}
{\lstset{language=c++}}
{}

\lstnewenvironment{bash}
{\lstset{language=Python}}
{}